\documentclass[apj]{emulateapj}

\newcommand{\kms}{\hbox{ km\thinspace s$^{-1}$}}    

\slugcomment{}

\shorttitle{Multiple Collimated Outflows in IC\,4634}
\shortauthors{Guerrero et al.}

\begin{document}


\title{Multiple and Precessing Collimated Outflows in the Planetary 
Nebula IC\,4634} 


\author{Mart\'{\i}n A.\ Guerrero\altaffilmark{1,2}}
 
\author{Luis F.\ Miranda\altaffilmark{1}}

\author{Angels Riera\altaffilmark{3,4}}

\author{Pablo F.\ Vel\'azquez\altaffilmark{5}}

\author{Lorenzo Olgu\'{\i}n\altaffilmark{6}}

\author{Roberto V\'azquez\altaffilmark{6}}

\author{You-Hua Chu\altaffilmark{7}}

\author{Alejandro Raga\altaffilmark{5}}

\author{G. Ben\'{\i}tez\altaffilmark{6}}



\altaffiltext{1}{Instituto de Astrof\'{\i}sica de Andaluc\'{\i}a, CSIC, 
       Apdo.\ Correos 3004, E-18080 Granada, Spain}
\altaffiltext{2}{Electronic address: mar@iaa.es}
\altaffiltext{3}{Departament de F\'{\i}sica i Enginyeria Nuclear, EUETIB,  
             Universitat Polit\`ecnica de Catalunya, 
             Compte d'Urgell 187, E-08036 Barcelona, 
             Spain}
\altaffiltext{4}{Departament d'Astronomia i Meteorologia, Universitat de 
             Barcelona, 
             Av. Diagonal 647, 08028 Barcelona, Spain}
\altaffiltext{5}{
Instituto de Ciencias Nucleares, Universidad Nacional Aut\'onoma de 
M\'exico, Ciudad Universitaria, 04510, Mexico City, Mexico}
\altaffiltext{6}{
Instituto de Astronom\'{\i}a, UNAM, Ensenada, Mexico}
\altaffiltext{7}{University of Illinois at Urbana-Champaign, Department of 
             Astronomy, 
             1002 W.\ Green St., Urbana, IL61801, USA}


\begin{abstract}
With its remarkable double-S shape, IC\,4634 is an archetype of 
point-symmetric planetary nebulae (PN).  
In this paper, we present a detailed study of this PN using archival 
\emph{HST} WFPC2 and ground-based narrow-band images to investigate 
its morphology, and long-slit spectroscopic observations to determine 
its kinematics and to derive its physical conditions and excitation.  
The data reveal new structural components, including a distant 
string of knots distributed along an arc-like feature 
40$^{\prime\prime}$--60$^{\prime\prime}$ from the center of the 
nebula, a skin of enhanced {[O~{\sc iii}]/H$\alpha$} ratio 
enveloping the inner shell and the double-S feature, and a 
triple-shell structure.  
The spatio-kinematical study also finds an equatorial component of the 
main nebula that is kinematically independent from the bright inner 
S-shaped arc.  
We have investigated in detail the bow shock-like features in IC\,4634 and 
found that their morphological, kinematical and emission properties are 
consistent with the interaction of a collimated outflow with surrounding 
material.  
Indeed, the morphology and kinematics of some of these features can 
be interpreted using a 3D numerical simulation of a collimated outflow 
precessing at a moderate, time-dependent velocity.  
Apparently, IC\,4634 has experienced several episodes of point-symmetric 
ejections oriented at different directions with the outer S-shaped feature 
being related to an earlier point-symmetric ejection and the outermost 
arc-like string of knots being the relic of an even much earlier 
point-symmetric ejection.  
There is tantalizing evidence that the action of these collimated 
outflows has also taken part in the shaping of the innermost shell 
and inner S-shaped arc of IC\,4634.  
\end{abstract}





\section{Introduction}

Planetary nebulae (PNe) represent the final phases of the stellar 
evolution of low- and intermediate-mass stars.  
The shape of a PN is thought to be determined by the interaction of the 
current fast, tenuous stellar wind \citep[e.g.,][]{PP91} of its central 
star with the previous slow, dense wind of the Asymptotic Giant Branch 
(AGB) phase.  
In the so-called Interacting Stellar Winds (ISW) model of PN formation 
\citep{KPF78}, the critical parameter is the density distribution of 
the AGB wind \citep{B87}: 
a spherical symmetric AGB wind will result in a spherical PN, 
while a polar density gradient in the AGB wind will result in 
an elliptical or bipolar PN, depending on the degree of the 
density gradient.

This simplified view of PN formation has been challenged by the 
overwhelming body of observations of PNe showing small-scale 
features \citep[e.g., NGC\,7662,][]{Petal04}, collimated bipolar 
outflows \citep[e.g., NGC\,7009,][]{FMS04}, and point-symmetric 
morphologies, including multiple point-symmetric bubbles \citep[e.g., 
Hen\,2-47 and M1-37,][]{S00} and point-symmetric collimated outflows 
\citep[e.g., NGC\,6884,][]{MGT99}.  
All these features are difficult to interpret in the framework of the ISW 
model and have inspired new theoretical ideas to explain the shaping of 
PNe.  
New scenarios of PNe formation and shaping include magnetic collimation 
\citep{G-SL00}, the action of collimated fast winds \citep{LS03}, or 
the blowing up of a warped circumstellar disk by the fast stellar wind 
\citep{RMI05}.

\citet{SCS93} recognized a new morphological class of PNe characterized 
by the point-symmetry with respect to their centers.  
The morphology of these PNe suggests that they are shaped by a mechanism 
involving precession, most likely the action of precessing or rotating 
collimated outflows.  
An investigation of PNe with point-symmetric morphologies is, thus, essential 
to evaluate the action of precessing collimated outflows in the formation of 
PNe and to assess the origin of the collimating mechanisms.  
Among the PNe with point-symmetric morphology, IC\,4634 is an outstanding 
case.  
IC\,4634 is an archetypical point-symmetric PN as \citet{S93} found 
that its remarkable double-S-shaped morphology is reflected in its 
kinematics: 
each component of the two S-shaped features expands radially with 
its counterpart on the other side of the star expanding in opposite 
direction.
The collimated outflows of IC\,4634 are further revealed by bow shock 
features indicating their interaction with the surrounding medium.  
Therefore, both the morphology and kinematics of IC\,4634 suggest the 
presence of highly symmetric fast collimated outflows, and the precession 
or rotation of the source that produced them.

Since the study of IC\,4634 by \citet{S93}, several studies have further 
investigated the kinematical and morphological properties of IC\,4634 
\citep{Hetal97,Tetal03}, but a detailed study is lacking.  
In this paper, we present a spatio-kinematical study of 
IC\,4634 using narrow-band archival \emph{HST} WFPC2 and 
ground-based images and complementary long-slit echelle 
and medium-dispersion spectroscopic observations.  
The observations, described in $\S$2, have allowed us to investigate 
the morphology, kinematics, and physical conditions of IC\,4634 and 
the possible shock excitation of the bow shock features.  
A thorough description of the different structural components 
of IC\,4634 derived from these observations is given in $\S$3, 
and the action of precessing collimated outflows is examined and 
modeled in $\S$4.  
The results are discussed in $\S$5.

\section{Observations}

\subsection{Narrow-Band Imaging Observations}

\begin{figure*}[!t]
\centerline{\includegraphics[width=15cm]{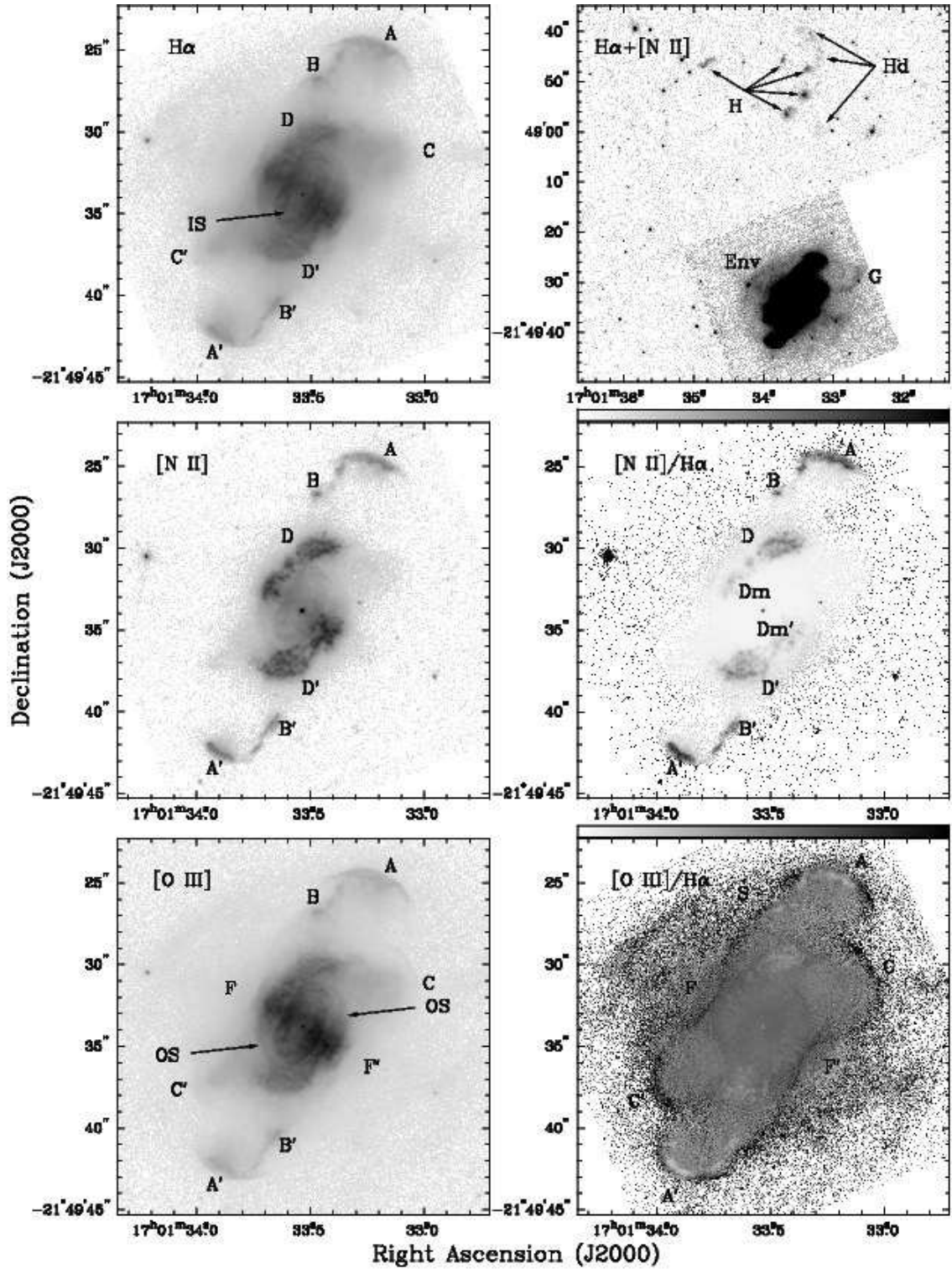}}
\caption{
{\it (left column)} \emph{HST} WFPC2 narrow-band images of the central 
region of IC\,4634 in the H$\alpha$ {\it (top)}, [N~{\sc ii}] {\it 
(center)}, and [O~{\sc iii}] {\it (bottom)} emission lines.  
{\it (right column)} \emph{HST} WFPC2 H$\alpha$+[N~{\sc ii}] image 
of IC\,4634 {\it (top)} showing the outermost emission found in 
this nebula, and [N~{\sc ii}]/H$\alpha$ {\it (center)}, and 
[O~{\sc iii}]/H$\alpha$ {\it (bottom}) ratio maps of the central 
region of IC\,4634.  
The different features identified in the nebula and described in the 
text are marked on these images.   
The images are displayed in a logarithmic scale, while the ratio 
maps are shown in linear scale.  
The grayscale upper limits of the ratio maps are 1.0 and 4.0 for 
the [N~{\sc ii}]/H$\alpha$ and [O~{\sc iii}]/H$\alpha$ ratio maps, 
respectively.  
}
\label{f1}
\end{figure*}

Narrow-band \emph{HST} WFPC2 images of IC\,4634 in the H$\beta$, H$\alpha$, 
[N~{\sc ii}] $\lambda$6584, and [O~{\sc iii}] $\lambda$5007 lines were 
retrieved from the \emph{HST} archive (Proposal ID~6856, PI: J.T.\ Trauger).  
The total integration times were 800 s for the H$\beta$ image and 
1,000 s for the H$\alpha$, [N~{\sc ii}], and [O~{\sc iii}] images.  
These images were calibrated via the pipeline procedure.  
Cosmic rays were removed by combining different exposures obtained 
with the same filter, using standard IRAF routines.  
The \emph{HST} WFPC2-PC H$\alpha$, [N~{\sc ii}], and [O~{\sc iii}] 
images of IC\,4634 are displayed in Figure~\ref{f1}-{\it left}.  
The H$\beta$ image, not shown here, has been used in conjunction 
with the H$\alpha$ image to investigate the amount of extinction 
towards IC\,4634 and its spatial distribution.  
The mosaiced \emph{HST} WFPC2 H$\alpha$ and [N~{\sc ii}] images 
have been added together to produce the deep H$\alpha$+[N~{\sc ii}] 
image shown in Fig.~\ref{f1}-{\it top-right}.

The \emph{HST} WFPC2-PC H$\alpha$, [N~{\sc ii}], and [O~{\sc iii}] 
images of IC\,4634 have been used to derive [N~{\sc ii}]/H$\alpha$, 
[O~{\sc iii}]/H$\alpha$, and [N~{\sc ii}]/[O~{\sc iii}] ratio maps.   
In Fig.~\ref{f1}-{\it right-center} and {\it -bottom}, the 
[N~{\sc ii}]/H$\alpha$ and [O~{\sc iii}]/H$\alpha$ 
ratio maps are shown, respectively.  
The [N~{\sc ii}]/[O~{\sc iii}] ratio map is not presented here because 
it is very similar to the [N~{\sc ii}]/H$\alpha$ ratio map.

Additional narrow-band images of IC\,4634 were obtained at the 2.5m 
Nordic Optical Telescope (NOT) of the Roque de los Muchachos Observatory 
(La Palma, Spain) on 2005 August 1.  
The images were acquired through filters that included the H$\alpha$ 
($\lambda_{\rm c}$=6563~\AA, $FWHM$=9~\AA, $t_{\rm exp}$=900~s) and 
H$\alpha$+[N~{\sc ii}] 
($\lambda_{\rm c}$=6562~\AA, $FWHM$=46~\AA, $t_{\rm exp}$=1800~s) 
emission lines using the ALFOSC (Andalucia Faint Object Spectrograph 
and Camera) camera in imaging mode.  
The detector was an E2V 2K$\times$2K CCD with a pixel size of 13.5$\mu$m, 
providing a plate scale of 0$\farcs$19~pixel$^{-1}$ and a field of view 
of $\sim$6$\farcm$5.  
The angular resolution during the observations, as derived from the 
$FWHM$ of stars in the field of view, was 0$\farcs$8--0$\farcs$9.

\subsection{Echelle Observations}

High-dispersion spectroscopic observations of the H$\alpha$ and 
[N~{\sc ii}] $\lambda\lambda$6548,6584 \AA\ lines of IC\,4634 were 
obtained using the echelle spectrograph on the 4m Blanco Telescope 
of the Cerro Tololo Inter-American Observatory (CTIO) on 2002 June 
22.
The spectrograph was used in the long-slit mode to obtain single-order 
observations of the H$\alpha$ and [N~{\sc ii}] $\lambda\lambda$6548,6584 
\AA\ lines.  
The unvignetted slit length is $\sim$3\arcmin.  
The 79 line~mm$^{-1}$ echelle grating and the long-focus red camera 
were used, resulting in a reciprocal dispersion of 3.4 \AA~mm$^{-1}$.  
The data were recorded with the SITe 2K No.\ 6 CCD with a pixel size of 
24 $\mu$m.  
This configuration provides a spatial scale of 0\farcs26 pixel$^{-1}$ 
and a sampling of 3.7\kms~pixel$^{-1}$ along the dispersion direction.  
The slit width was set to 0\farcs9, and the resultant instrumental 
resolution ($FWHM$) was 8\kms.  
The angular resolution, determined by the seeing, was $\sim$1\farcs0.  
The echelle observations were made with the slit placed at the 
central star of IC\,4634 and oriented along the position angles (PA) 
60$^\circ$ ($t_{\rm exp}$=1050~s), 
120$^\circ$ and 180$^\circ$ ($t_{\rm exp}$=900~s), and 
150$^\circ$ and 166$^\circ$ ($t_{\rm exp}$=1200~s).

High-dispersion spectroscopic observations of the H$\alpha$, 
[N~{\sc ii}] $\lambda$6584 \AA, and [O~{\sc iii}] $\lambda$5007 
\AA\ lines were also obtained using the IACUB spectrograph 
\citep{MKetal93} on the 2.5m NOT of the Roque de los Muchachos 
Observatory (La Palma, Spain) on 2004 June 29.  
The spectrograph was used in the long-slit mode to obtain single-order 
observations of the H$\alpha$ and [N~{\sc ii}] $\lambda$6584 \AA\ lines 
in the 9$^{\rm th}$ order, and of the [O~{\sc iii}] $\lambda$5007 \AA\ 
line in the 11$^{\rm th}$ order.  
IACUB provides a reciprocal dispersion of 1.74 \AA~mm$^{-1}$ in 
the 9$^{\rm th}$ order and 1.43  \AA~mm$^{-1}$ in the 11$^{\rm th}$ 
order.  
The data were recorded with a Thomson CCD with a pixel size of 
19 $\mu$m.  
This configuration provides a spatial scale of 0\farcs139 pixel$^{-1}$, 
and a sampling along the dispersion direction of 1.5\kms~pixel$^{-1}$ 
for the spectrum of the H$\alpha$ and [N~{\sc ii}] $\lambda$6584 \AA\ 
lines, and 1.6\kms~pixel$^{-1}$ for the spectrum of the [O~{\sc iii}] 
$\lambda$5007 \AA\ line.  
The unvignetted slit length is 40\arcsec\ and the slit width was 
set to 0\farcs65, resulting in an instrumental resolution ($FWHM$) 
of 9.5\kms\ 
for the H$\alpha$ and [N~{\sc ii}] $\lambda$6584 \AA\ lines, and 
8.5\kms\ for the [O~{\sc iii}] $\lambda$5007 \AA\ line.  
The angular resolution, determined by the seeing, was $\sim$0\farcs8.  
The echelle observations of the H$\alpha$ and [N~{\sc ii}] $\lambda$6584 
\AA\ lines was made along a slit offset 2\arcsec\ West and 46\farcs4 North 
of the central star of IC\,4634 and oriented along PA=0$^\circ$ 
($t_{\rm exp}$=1200 s).  
The echelle observations of the [O~{\sc iii}] $\lambda$5007 \AA\ 
line was made along a slit placed at the central star of IC\,4634 
and oriented along PA=150$^\circ$ ($t_{\rm exp}$=1800 s).

\begin{deluxetable}{ccrrc}[b!]
\tablewidth{0pt}
\tabletypesize{\scriptsize}
\tablecaption{Log of the Medium-Dispersion Spectroscopic Observations}
\tablehead{
\colhead{Date}           & 
\colhead{Slit Position}  &
\colhead{Position Angle} & 
\colhead{$t_{\rm exp}$}  & 
\colhead{Number}         \\
\colhead{}               & 
\colhead{}               & 
\colhead{[arcdeg]}       & 
\colhead{[s]}            & 
\colhead{of Spectra}     }
\startdata
2002 June 16 & Central Star  &   153~~~~~~~~~ &       900     &   1    \\ 
             & Central Star  &   153~~~~~~~~~ &       300     &   4    \\
             & Central Star  &   153~~~~~~~~~ &        60     &   3    \\ 
             & Central Star  &   153~~~~~~~~~ &        20     &   3    \\
\hline                                                  
2004 May 24  & Central Star  &   153~~~~~~~~~ &       300     &   3    \\ 
             & Central Star  &   153~~~~~~~~~ &        10     &   3    \\ 
\hline                                                  
2004 August  & Central Star  &    90~~~~~~~~~ &       120     &   3    \\
             & Central Star  &    90~~~~~~~~~ &        30     &   1    \\ 
             & Central Star  &    90~~~~~~~~~ &        20     &   1    \\ 
             & Central Star  &    90~~~~~~~~~ &        10     &   4    \\ 
             & A~            &    90~~~~~~~~~ &       300     &   3    \\
             & A$^\prime$    &    90~~~~~~~~~ &       600     &   3    \\
             & D~            &    90~~~~~~~~~ &        30     &   4    \\
             & D$^\prime$    &    90~~~~~~~~~ &       300     &   3    
\enddata
\end{deluxetable}

\begin{deluxetable*}{lrrrrrrrrrr}[t!]
\tabletypesize{\scriptsize}
\tablecaption{Measured and Intrinsic Line Intensity Ratios for IC\,4634}
\tablewidth{0pt}
\tablehead{
\multicolumn{1}{c}{}              & 
\multicolumn{2}{c}{\underline{~~~~~~~IS~~~~~~~~~~}} & 
\multicolumn{2}{c}{\underline{~~~~~~~A~~~~~~~~~~}} & 
\multicolumn{2}{c}{\underline{~~~~~~~A$^\prime$~~~~~~~~~~}} & 
\multicolumn{2}{c}{\underline{~~~~~~~D~~~~~~~~~~}} & 
\multicolumn{2}{c}{\underline{~~~~~~~D$^\prime$~~~~~~~~~~}} \\
\multicolumn{1}{l}{Line}          & 
\multicolumn{1}{c}{$F$}           & 
\multicolumn{1}{c}{$I$}        & 
\multicolumn{1}{c}{$F$}        & 
\multicolumn{1}{c}{$I$}        & 
\multicolumn{1}{c}{$F$}        & 
\multicolumn{1}{c}{$I$}        & 
\multicolumn{1}{c}{$F$}        & 
\multicolumn{1}{c}{$I$}        & 
\multicolumn{1}{c}{$F$}        & 
\multicolumn{1}{c}{$I$}        }
\startdata
~H~{\sc i}      $\lambda$4340 &  40$\pm$2       &  44~~~ &  41$\pm$3     &   45~~~ &  42$\pm$5      &   46~~~ &  41$\pm$3     &   45~~~ &  37$\pm$6     &   41~~~ \\  
~[O~{\sc iii}]  $\lambda$4363 &   5.7$\pm$0.3   &    6.3 &   7.5$\pm$1.6 &     8.3 &   6.0$\pm$0.8  &     6.6 &   6.0$\pm$0.4 &     6.6 &   5.0$\pm$0.9 &     5.5 \\  
~He~{\sc i}     $\lambda$4387 &   1.2$\pm$0.2   &    1.3 &   $\dots$     & $\dots$ &   $\dots$      & $\dots$ &   $\dots$     & $\dots$ &   $\dots$     & $\dots$ \\  
~He~{\sc i}     $\lambda$4471 &   5.2$\pm$0.3   &    5.6 &   5.4$\pm$1.2 &     5.8 &   5.2$\pm$0.7  &     5.6 &   5.2$\pm$0.4 &     5.6 &   4.6$\pm$0.6 &     5.0 \\  
~N~{\sc iii}$\lambda$4641 + O~{\sc ii}$\lambda$4649 & & & & & & & & & & \\
~~~~~~~~~~~~~~~+ C~{\sc iv}$\lambda$4658 
                              &   1.1$\pm$0.2   &    1.1 &   $\dots$     & $\dots$ &   $\dots$      & $\dots$ &   $\dots$     & $\dots$ &   $\dots$     & $\dots$ \\
~He~{\sc ii}    $\lambda$4686 &   0.5$\pm$0.2   &    0.5 &   $\dots$     & $\dots$ &   $\dots$      & $\dots$ &   $\dots$     & $\dots$ &   $\dots$     & $\dots$ \\
~[Ar~{\sc iv}]  $\lambda$4714 &   1.5$\pm$0.1   &    1.5 &   $\dots$     & $\dots$ &   $\dots$      & $\dots$ &   1.0$\pm$0.2 &     1.0 &   1.2$\pm$0.4 &     1.2 \\  
~[Ar~{\sc iv}]  $\lambda$4740 &   1.2$\pm$0.1   &    1.2 &   $\dots$     & $\dots$ &   $\dots$      & $\dots$ &   0.5$\pm$0.2 &     0.5 &   0.8$\pm$0.3 &     0.8 \\  
~H~{\sc i}      $\lambda$4861 & 100             & 100~~~ & 100           &  100~~~ & 100            &  100~~~ & 100           &  100~~~ & 100           &  100~~~ \\ 
~He~{\sc i}     $\lambda$4920 &   1.7$\pm$0.2   &    1.7 &   2.0$\pm$0.6 &     2.0 &   1.8$\pm$0.6  &     1.8 &   1.9$\pm$0.3 &     1.9 &   1.8$\pm$0.5 &     1.8 \\  
~[O~{\sc iii}]  $\lambda$4959 & 330$\pm$10      & 322~~~ & 291$\pm$8     &  284~~~ & 260$\pm$ 20    &  254~~~ & 290$\pm$ 15   &  283~~~ & 285$\pm$20    &  278~~~ \\ 
~[O~{\sc iii}]  $\lambda$5007 & 995$\pm$ 30     & 963~~~ & 876$\pm$18    &  847~~~ & 850$\pm$ 25    &  822~~~ & 870$\pm$ 30   &  841~~~ & 835$\pm$30    &  808~~~ \\ 
~[N~{\sc i}]    $\lambda$5199 &   0.14$\pm$0.05 &    0.13&   $\dots$     & $\dots$ &    $\dots$     & $\dots$ &   $\dots$     & $\dots$ &   1.7$\pm$0.5 &     1.6 \\  
~[Fe~{\sc iii}] $\lambda$5270 &   0.15$\pm$0.06 &    0.14&   $\dots$     & $\dots$ &    0.3$\pm$0.1 &     0.3 &   $\dots$     & $\dots$ &   $\dots$     & $\dots$ \\
~[Cl~{\sc iii}] $\lambda$5517 &   0.40$\pm$0.06 &    0.35&   0.7$\pm$0.2 &     0.6 &    0.6$\pm$0.2 &     0.5 &   0.6$\pm$0.2 &     0.5 &   0.4$\pm$0.2 &     0.4 \\   
~[Cl~{\sc iii}] $\lambda$5535 &   0.58$\pm$0.05 &    0.51&   0.8$\pm$0.3 &     0.7 &    0.6$\pm$0.2 &     0.5 &   0.7$\pm$0.2 &     0.6 &   0.6$\pm$0.2 &     0.5 \\   
~[N~{\sc ii}]   $\lambda$5755 &   0.35$\pm$0.04 &    0.30&   1.5$\pm$0.6 &     1.3 &    2.2$\pm$0.7 &     1.9 &   1.2$\pm$0.3 &     1.0 &   1.1$\pm$0.4 &     0.9 \\  
~C~{\sc iv}  $\lambda$5801-12 &   0.55$\pm$0.06 &    0.46&   $\dots$     & $\dots$ &    $\dots$     & $\dots$ &   $\dots$     & $\dots$ &   $\dots$     & $\dots$ \\
~He~{\sc i}     $\lambda$5876 &  17$\pm$1       &   14.1 &  16.9$\pm$1.0 &    14.0 &   16.2$\pm$1.2 &    13.4 &  17.0$\pm$1.5 &    14.1 &  19.7$\pm$1.3 &    16.4 \\  
~[O~{\sc i}]    $\lambda$6300 &   0.8$\pm$0.2   &    0.6 &   9.6$\pm$0.5 &     7.4 &    9.6$\pm$0.5 &     7.5 &   4.4$\pm$0.5 &     3.4 &   4.9$\pm$0.7 &     3.8 \\  
~[S~{\sc iii}]  $\lambda$6312 &   1.3$\pm$0.15  &    1.0 &   2.3$\pm$0.5 &     1.8 &    2.7$\pm$0.4 &     2.1 &   1.9$\pm$0.4 &     1.5 &   2.3$\pm$0.6 &     1.8 \\  
~[O~{\sc i}]    $\lambda$6363 &   0.30$\pm$0.10 &    0.23&   3.5$\pm$0.4 &     2.7 &    3.3$\pm$0.4 &     2.5 &   1.7$\pm$0.4 &     1.3 &   1.7$\pm$0.5 &     1.3 \\  
~[N~{\sc ii}]   $\lambda$6548 &   4.2$\pm$0.4   &    3.2 &  37$\pm$4     &   28~~~ &   42$\pm$4     &   31~~~ &  20$\pm$2     &    15.0 &  22$\pm$5     &    16.5 \\
~H~{\sc i}      $\lambda$6563 & 355$\pm$20      & 264~~~ & 379$\pm$10    &  283~~~ &  365$\pm$10    &  272~~~ & 390$\pm$10    &  290~~~ & 373$\pm$8     &  278~~~ \\
~[N~{\sc ii}]   $\lambda$6584 &  11.2$\pm$0.8   &    8.4 & 110$\pm$6     &   82~~~ &  107$\pm$6     &   80~~~ &  56$\pm$5     &   42~~~ &  65$\pm$6     &   49~~~ \\
~He~{\sc i}     $\lambda$6678 &   5.2$\pm$0.4   &    3.8 &   5.9$\pm$0.5 &     4.3 &    5.1$\pm$0.5 &     3.8 &   5.0$\pm$0.3 &     3.7 &   5.8$\pm$0.5 &     4.3 \\  
~[S~{\sc ii}]   $\lambda$6717 &   0.70$\pm$0.15 &    0.5 &  11.2$\pm$1.5 &     8.2 &   10.7$\pm$1.5 &     7.8 &   4.6$\pm$1.0 &     3.4 &   4.8$\pm$1.1 &     3.5 \\  
~[S~{\sc ii}]   $\lambda$6731 &   1.4$\pm$0.2   &    1.0 &  17.5$\pm$1.5 &    12.7 &   14.0$\pm$1.5 &    10.2 &   7.8$\pm$1.2 &     5.7 &   8.7$\pm$1.2 &     6.4 \\  
~He~{\sc i}     $\lambda$7065 &   8.0$\pm$0.7   &    5.6 &   5.3$\pm$0.3 &     3.7 &    5.2$\pm$0.4 &     3.6 &   5.5$\pm$0.6 &     3.9 &   7.2$\pm$0.7 &     5.1 \\  
~[Ar~{\sc iii}] $\lambda$7135 &  13.5$\pm$0.8   &    9.4 &  18.1$\pm$1.0 &    12.5 &   16.5$\pm$1.1 &    11.5 &  18.3$\pm$1.3 &    12.7 &  20.2$\pm$1.5 &    14.0 \\  
~He~{\sc i}     $\lambda$7281 &   1.1$\pm$0.2   &    0.8 &   $\dots$     & $\dots$ &    $\dots$     & $\dots$ &   1.1$\pm$0.2 &     0.8 &   1.1$\pm$0.2 &     0.8 \\  
~[O~{\sc ii}]   $\lambda$7320 &   3.0$\pm$0.3   &    2.0 &   9.4$\pm$0.8 &     6.3 &    9.0$\pm$0.7 &     6.1 &   $\dots$     & $\dots$ &   9.8$\pm$0.7 &     6.7 \\  
~[O~{\sc ii}]   $\lambda$7330 &   2.8$\pm$0.3   &    1.9 &   7.4$\pm$0.9 &     5.0 &    8.1$\pm$0.8 &     5.5 &   $\dots$     & $\dots$ &   8.2$\pm$0.8 &     5.6 \\  
                              &                 &        &               &         &                &         &               &         &               &         \\
~F(H$\beta$) 
    (ergs~cm$^{-2}$~s$^{-1}$) & 1.4$\times$10$^{-11}$ &  & 4.0$\times$10$^{-12}$ & & 3.5$\times$10$^{-12}$ &  & 2.6$\times$10$^{-12}$ & & 2.1$\times$10$^{-12}$ & 
\enddata
\end{deluxetable*}

\subsection{Medium-Dispersion Spectroscopic Observations}

Medium-dispersion long-slit spectroscopic observations of IC\,4634 were 
obtained using the Boller \& Chivens spectrograph on the 2.1m telescope 
of the Observatorio Astron\'omico Nacional de San Pedro M\'artir (Baja 
California, Mexico) on 2002 June 16 and 2004 May 20 and August 12.  
Multiple observations were obtained at several slit positions 
as listed in Table~1.  
In all cases, the 400 line~mm$^{-1}$ grating was used, the slit length 
was 5$^\prime$,  and its width was set to 150~$\mu$m, projecting to 
2\arcsec\ on the sky.  
The data were recorded on a SITe3 1K CCD with pixel size 
of 24$\mu$m.  
This configuration provides a spatial scale of 1\farcs05 pixel$^{-1}$ 
and a spectral scale of 3 \AA~pixel$^{-1}$ with a spectral coverage 
from 4240 \AA\ to 7310 \AA.  
The long-slit spectra were reduced and calibrated following standard
procedures using the XVISTA package\footnote{
XVISTA was originally developed as Lick Observatory Vista.  
It is currently maintained by Jon Holtzman at New Mexico State 
University and is available at 
http://ganymede.nmsu.edu/holtz/xvista.}
.
For the wavelength calibration, we used a He-Ar lamp.  
The spectral resolution was $\sim$6.8 \AA.  
For the flux calibration, several standard stars were observed 
each night.

\section{Structural Components of IC\,4634}

The \emph{HST} narrow-band images and ratio maps of IC\,4634 shown in 
Fig.~\ref{f1} reveal a wealth of structural components in this nebula.  
In Fig.~\ref{f1}, we have labeled the most prominent structures in IC\,4634:  
\begin{itemize}
\item 
a bright inner shell (\emph{IS}), surrounded by a dim outer shell (OS), 
\item 
an inner S-shaped feature formed by two knotty point-symmetric arcs 
(\emph{D-D$^\prime$}) that present distinct properties along the minor 
axis of \emph{IS} (\emph{Dm-Dm$^\prime$}), 
\item 
an outer S-shaped feature formed by two pairs of bow shock-like 
structures (\emph{A-A$^\prime$} and \emph{B-B$^\prime$}) along 
different PAs and located at different distances from the central 
star,  
\item
a pair of bow shock-like structures, \emph{C-C$^\prime$}, along 
PA$\sim$120\arcdeg, 
\item 
a skin of enhanced [O~{\sc iii}]/H$\alpha$ ratio (\emph{S}) that 
encloses the inner shell and bow shock-like structures, 
\item 
an outer envelope (\emph{Env}) surrounding all previous components 
and where some individual features (\emph{F-F$^\prime$} and \emph{G}) 
can be distinguished, 
\item 
a string of knots (\emph{H}) distributed along an arc-like feature 
at $\sim$40$^{\prime\prime}$ North of IC\,4634 and accompanied by 
diffuse emission (\emph{Hd}).  
\end{itemize}
In the following, we will describe in greater detail the 
morphology, kinematics, excitation, and physical conditions 
of the different structural components present in IC\,4634.

\begin{deluxetable*}{lrrrrr}{t!}
\tablewidth{0pt}
\tablecaption{Plasma Diagnostics and Ionic and Elemental Abundances for IC\,4634}
\tablehead{
\colhead{}           & 
\colhead{IS} & 
\colhead{A}  & 
\colhead{A$^\prime$} & 
\colhead{D}  & 
\colhead{D$^\prime$} }
\startdata
Plasma Diagnostics                   &        & & & & \\
$N_{\rm e}$ [S~{\sc ii}] (cm$^{-3}$) & 10,600 &  2,800 &  1,500 &  3,800 &  5,200 \\
$T_{\rm e}$ [N~{\sc ii}] (K)         & 13,150 &  9,600 & 11,600 & 11,550 & 10,250 \\
$T_{\rm e}$ [O~{\sc iii}] (K)        &  9,850 & 11,350 & 10,700 & 10,550 & 10,050 \\
\hline
Ionic Abundances                     &        &        &        &        &        \\
He$^{+}$/H$^+$                       &  0.096 &  0.097 &  0.098 &  0.098 &  0.113 \\
He$^{++}$/H$^+$                      &  0.001 &  0.000 &  0.000 &  0.000 &  0.000 \\
O$^0$/H$^+$ ($\times$10$^7$)         &  6.2   &  150   &  95    &  44    &  70    \\
O$^+$/H$^+$ ($\times$10$^6$)         &  7.1   &  110   &  60    &$\dots$ &  77    \\
O$^{++}$/H$^+$ ($\times$10$^4$)      &  3.7   &  2.0   &  2.4   &  2.6   &  2.8   \\
N$^+$/H$^+$ ($\times$10$^6$)         &  1.3   &  16    &  12    &  6.6   &  9.9   \\
S$^+$/H$^+$ ($\times$10$^8$)         &  7.1   &  70    &  41    &  27    &  43    \\
Cl$^{++}$/H$^+$ ($\times$10$^8$)     &  6.5   &  6.6   &  6.2   &  6.7   &  6.0   \\
Ar$^{+3}$/H$^+$ ($\times$10$^7$)     &  2.1   &  8.7   &  9.2   &  10    &  13    \\
\hline
Elemental Abundances                 &        &        &        &        &        \\
He/H                                 & 0.097  & 0.097  & 0.098  & 0.098  & 0.113  \\
O/H ($\times$10$^4$)                 & 3.7    & 3.1    & 3.0    & 2.6    & 3.6    \\
N/H ($\times$10$^4$)                 & 0.67   & 0.47   & 0.62   &$\dots$ & 0.46   \\
S/H ($\times$10$^4$)                 & 0.025  & 0.059  & 0.043  &$\dots$ & 0.053  \\
Cl/H ($\times$10$^4$)                & 0.0007 & 0.0007 & 0.0006 & 0.0007 & 0.0006 \\
Ar/H ($\times$10$^4$)                & 0.011  & 0.009  & 0.012  & 0.021  & 0.018  
\enddata
\end{deluxetable*}

\begin{figure*}[!t]
\centerline{\includegraphics[width=17.5cm]{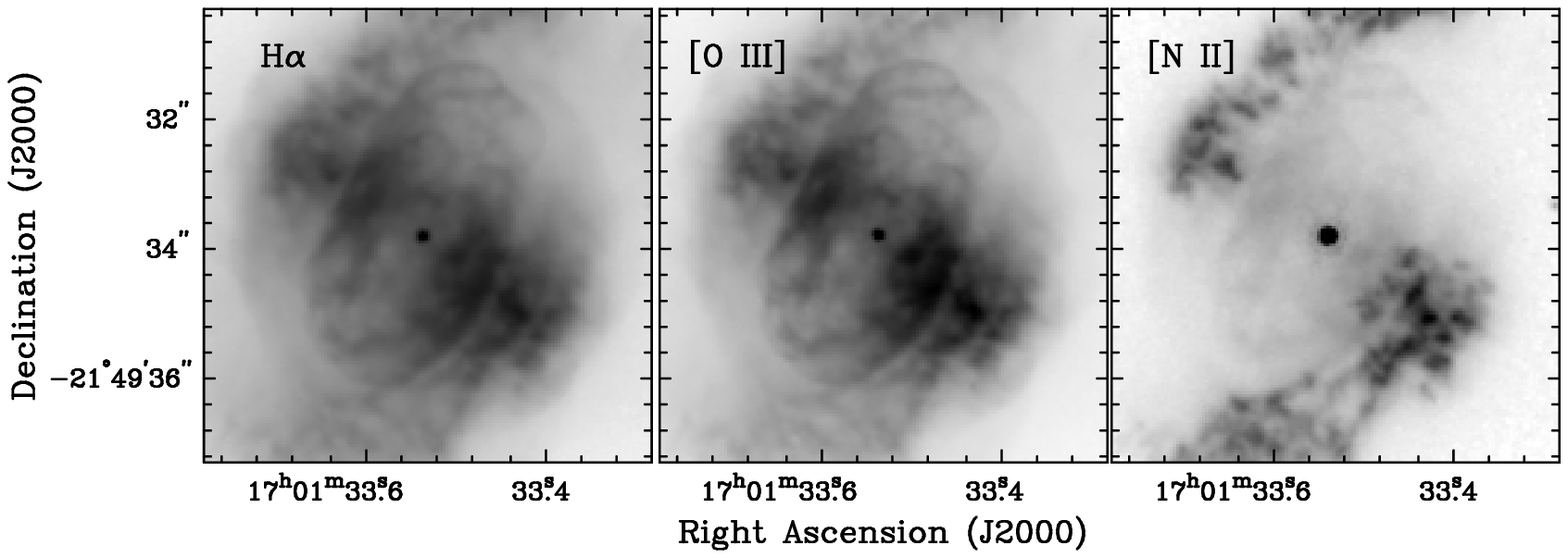}}
\caption{
\emph{HST} WFPC2 narrow-band images of the inner shell of IC\,4634 in the 
H$\alpha$ {\it (left)}, [O~{\sc iii}] {\it (center)}, and [N~{\sc ii}] 
{\it (right)} emission lines.  
}
\label{f2}
\end{figure*}

The H$\beta$ fluxes, and measured and intrinsic line intensity 
ratios of the emission lines detected in the medium-dispersion 
spectra in regions \emph{IS}, \emph{A-A$^\prime$}, and 
\emph{D-D$^\prime$} are listed in Table~2.  
The H$\alpha$ to H$\beta$ ratio measured in the medium dispersion 
spectra and in the \emph{HST} images imply an extinction coefficient, 
$c_{{\rm H}\beta}=0.34$.  
The \emph{HST} H$\alpha$/H$\beta$ ratio map indicates that there are 
no significant extinction variations across IC\,4634.  
The measured line intensity ratios have been dereddened accordingly 
using the wavelength-dependent extinction law from \citet{SM79}.  
The plasma diagnostics ($T_{\rm e}$ and $N_{\rm e}$), and ionic and 
elemental abundances listed in Table~3 were derived from the intrinsic 
line intensity ratios listed in Tab.~2 for the different regions of 
IC\,4634 using the nebular abundance package ELSA \citep{Jetal06}.

\subsection{The Inner Shells of IC\,4634}

From the center outward, the first component of IC\,4634 is the inner 
shell, \emph{IS}, around the central star.  
This inner shell has a size of 2\farcs5$\times$5\farcs2 and a major 
axis along PA$\sim$150$^\circ$.  
The edge of the inner shell is sharp and its surface brightness is 
enhanced with respect to the shell interior.  
This morphology suggests a thin shell, although it must be 
noted that an intricate pattern of filaments is superimposed 
on the shell.  
The shell brightness is also notably enhanced along its minor 
axis, marking a bright equatorial belt.  
The shape of the inner shell can be roughly described as elliptical 
(Figure~\ref{f2}), but, at the tips of its major axis, several protrusions 
of different sizes extend outwards.  
Further distortion of the shell shape is introduced by the bending 
of the shell edge at its equator, making the edge of the inner shell 
along its major axis to show a subtle S-shape.  
It is worthwhile to note that the central star of IC\,4634 is 
not located at the exact center of the shell, but displaced by 
$\sim$0\farcs20 towards the South.

\begin{figure*}[!t]
\centerline{\includegraphics[width=17.5cm]{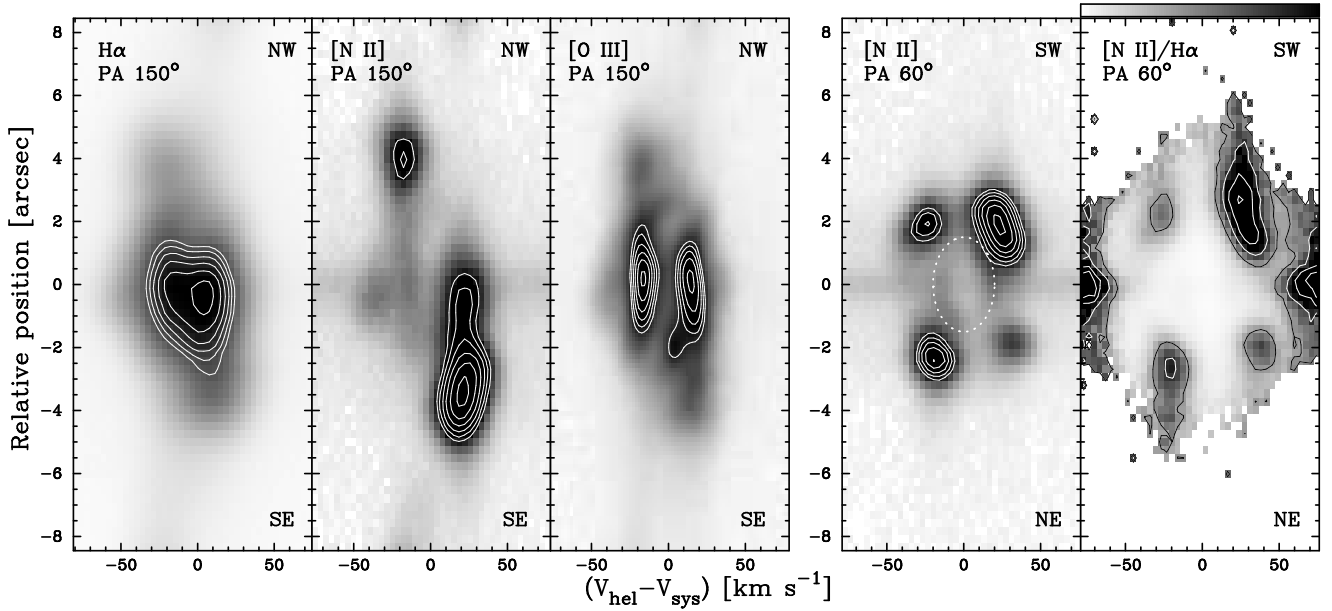}}
\caption{
Echellograms of the H$\alpha$, [N~{\sc ii}] $\lambda$6584 \AA, 
and [O~{\sc iii}] $\lambda$5007 \AA\ emission lines along 
PA 150$^\circ$, and echellogram of [N~{\sc ii}] $\lambda$6584 
\AA\ emission line and [N~{\sc ii}]/H$\alpha$ line ratio map 
along PA 60$^\circ$ of the inner shell of IC\,4634.  
In the [N~{\sc ii}] echellograms, the emission is dominated by the 
adjacent regions \emph{D-D$^\prime$} and \emph{Dm-Dm$^\prime$}, 
but still a hollow cavity can be distinguished, as marked by an 
ellipse in the [N~{\sc ii}] echellogram along PA 60$^\circ$.  
The echellograms are displayed in a logarithmic scale, while the 
line ratio map is shown in linear scale with lower limit 0 and 
upper limit 0.25.  
}
\label{f3}
\end{figure*}

The echellogram of the [O~{\sc iii}] emission line along the major 
axis of the inner shell at  PA~150\arcdeg\ (Figure~\ref{f3}) reveals a 
prolate expanding shell with an expansion velocity at the position 
of the central star of 15.3 km~s$^{-1}$.  
The systemic radial velocity in the Local Standard of Rest (LSR) 
system is --20.1 km~s$^{-1}$, in agreement with other estimates 
\citep[e.g.,][]{DAZ98}.  
The line tilt is small, indicating a low inclination of the shell 
with respect to the plane of the sky, but the S-shaped pattern 
of the line suggests accelerating features at the tips of the 
shell major axis.  
These blowout structures correspond to the protrusions observed at 
the tips of the major axis of the shell, with the Northwest protrusion 
receding from the observer and the Southeast protrusion approaching.  
The echellogram of the [N~{\sc ii}] emission line along the major 
(PA=150$^\circ$) and minor (PA=60$^\circ$) axes (Fig.~\ref{f3}) also 
reveal an expanding shell, but its kinematics cannot be studied 
in detail in this line because the bright emission from the adjacent 
\emph{D-D$^\prime$} and \emph{Dm-Dm$^\prime$} regions overwhelms the 
emission from \emph{IS}.   
The expansion velocity at the position of the central star is 
illustrated in the [N~{\sc ii}] echellogram along the minor axis 
by an ellipse with spatial semi-axis $\sim$1\farcs3 and expansion 
velocity of 18 km~s$^{-1}$.  
In the H$\alpha$ echellogram along the major axis of IC\,4634, also 
shown in Fig.~\ref{f3}, the thermal broadening of the line is larger than 
the line split and no shell can be clearly identified.

\begin{figure*}[t!]
\centerline{\includegraphics[width=17.5cm]{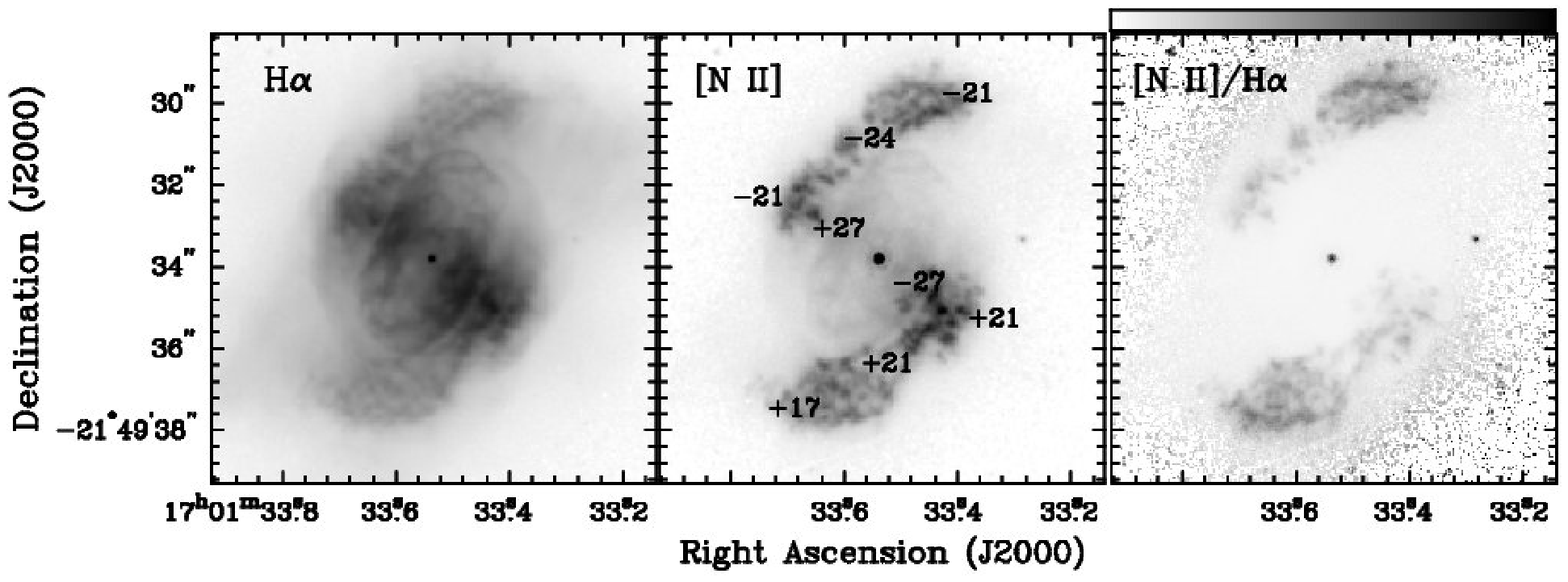}}
\caption{
\emph{HST} WFPC2 narrow-band images of IC\,4634 in the H$\alpha$ 
{\it (left)} and [N~{\sc ii}] {\it (center)} emission lines, and 
[N~{\sc ii}]/H$\alpha$ ratio map {\it (right)} emphasizing the 
emission from regions \emph{D-D$^\prime$} and \emph{Dm-Dm$^\prime$}.  
The radial velocities with respect to the systemic velocity 
along the \emph{D-D$^\prime$} arc-like features and at the 
location of \emph{Dm-Dm$^\prime$} are marked on the [N~{\sc ii}] 
image.  
The images and ratio maps are presented in a linear scale.  
The greyscale of the ratio map has a lower limit of 0 and and an 
upper limit of 1.  
}
\label{f4}
\end{figure*}

\begin{figure*}[!t]
\centerline{
\includegraphics[height=9cm]{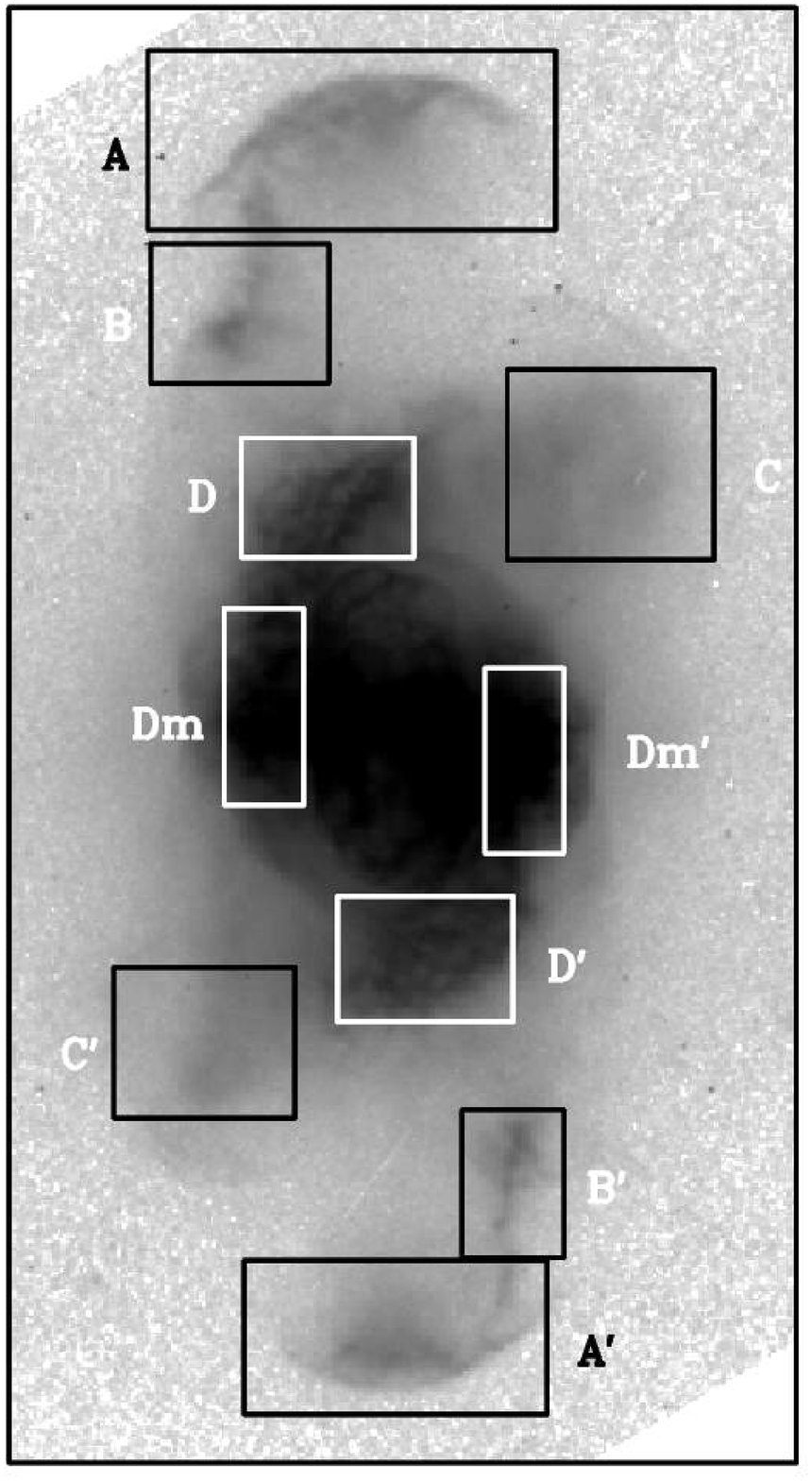}
\includegraphics[height=9.1cm]{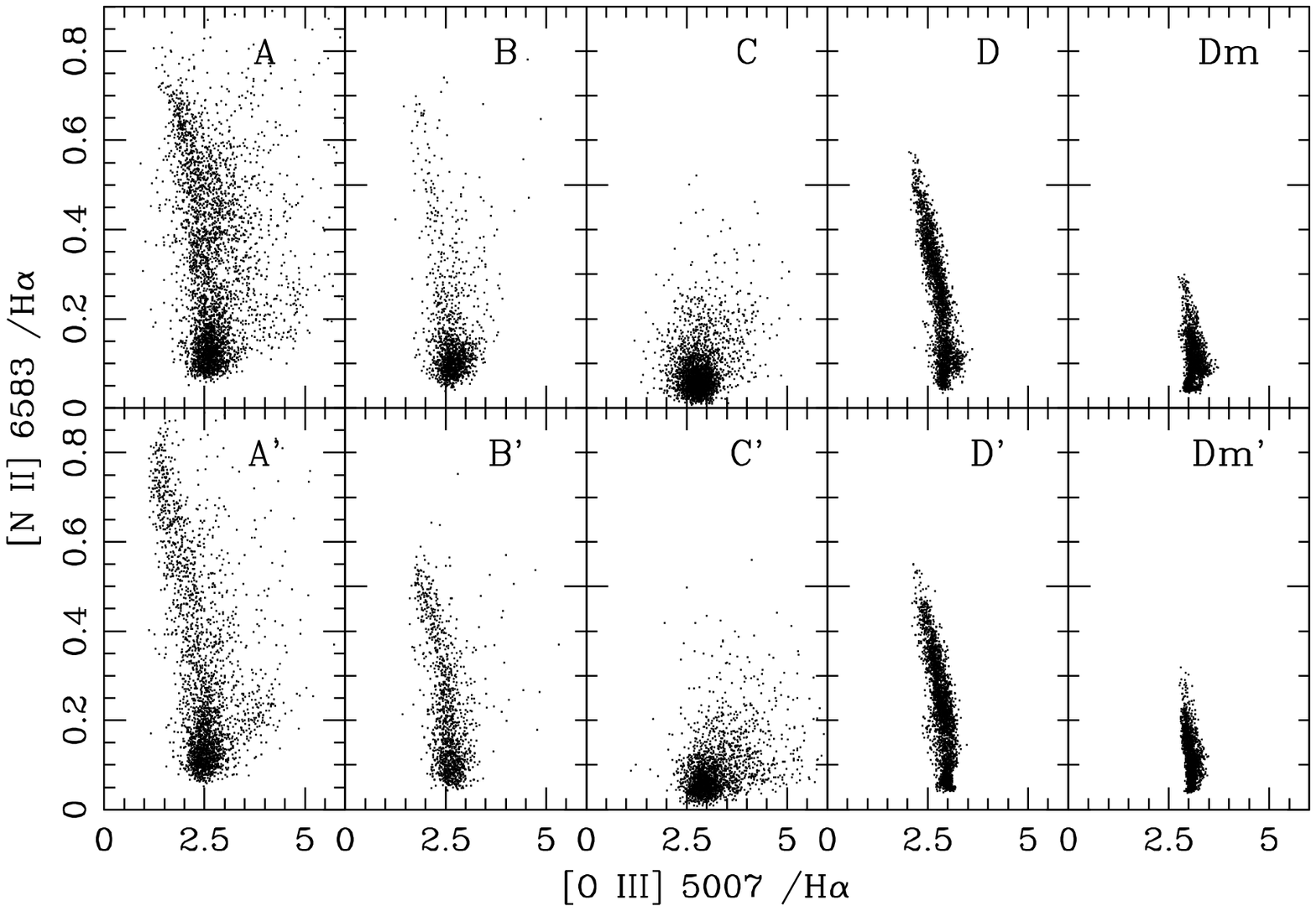}
}
\caption{
{\it (left)} [O~{\sc iii}]/H$\alpha$ vs.\ [N~{\sc ii}]/H$\alpha$ 
diagnostic diagrams of different morphological components of 
IC\,4634.  
The exact spatial regions from which these diagnostic diagrams have 
been extracted are overlaid on an H$\alpha$ image of IC\,4634 {\it 
(right)}.  
}
\label{f5}
\end{figure*}

The inner shell of IC\,4634 is the region where the 
[N~{\sc ii}]/H$\alpha$ ratio map (Fig.~\ref{f1}) shows its 
lowest values, $\sim$0.04.  
The high excitation is confirmed by the [N~{\sc ii}]/H$\alpha$ line 
ratio map presented in Fig.~\ref{f3} that shows this ratio to be nearly 
zero in the central regions of IC\,4634.  
The [O~{\sc iii}]/H$\alpha$ ratio (Fig.~\ref{f1}) is very flat in this 
region, with a typical dereddened value of $\sim$3.0.  
This is also the only region where emission lines of He~{\sc ii} are 
detected in the low-dispersion spectra (Tab.~1), although the intensities of 
these lines are low with $I$(He~{\sc ii} 4686 \AA)/$I$(H$\beta$)~$\sim0.005$.  
Besides the [Ar~{\sc iv}] $\lambda\lambda$4715,4740 \AA\ 
lines, no other high-excitation lines are present in the 
spectrum of the inner shell of IC\,4634.  
The [Fe~{\sc iii}] $\lambda$5270 \AA\ emission line and the Wolf-Rayet 
complex of C~{\sc iv} $\lambda$5801-12 \AA\ are weakly detected in this 
spectrum.

The electron density and temperature of the inner shell of IC\,4634 can be 
derived using the density sensitive [S~{\sc ii}] $\lambda\lambda$6716,6731 
\AA\ doublet line ratio and the temperature sensitive [N~{\sc ii}] and 
[O~{\sc iii}] auroral to nebular line ratios.  
Using the line intensities listed in Tab.~2, we have computed an 
electron density of $\sim$10,600~cm$^{-3}$ from the [S~{\sc ii}] 
line ratios, and electron temperatures of 13,000~K from the 
[N~{\sc ii}] lines and of 10,000~K from the [O~{\sc iii}] lines.  
These values are consistent with previous estimates of the physical 
conditions in the central region of IC\,4634 
\citep{AC83,dFPetal92,HAF99} that also find higher electron temperature 
from the [N~{\sc ii}] lines than from the [O~{\sc iii}] lines.  
The nebula does not show any evidence of N or He enrichment, 
thus it can be classified as a type~II PN based on its chemical 
abundances \citep{P78}.

The inner shell of IC\,4634 is surrounded by a faint shell of 
radius 3$^{\prime\prime}$ best seen at PAs $85^\circ$--150$^\circ$ 
and $265^\circ$--330$^\circ$ (Fig.~\ref{f2}), the outer shell \emph{OS}, 
as marked in Fig.~\ref{f1}.  
At other PAs, the low surface brightness of this shell is overwhelmed 
by the bright \emph{D-D$^\prime$} and \emph{Dm-Dm$^\prime$} features.  
The visible portion of \emph{OS} suggests that it is round 
or, at least, less aspherical than the innermost \emph{IS}.

\subsection{The Inner Point-Symmetric Structure}

One of the most prominent features in IC\,4634 is the pair of 
low-excitation arcs that we named \emph{D-D$^\prime$} in Fig.~\ref{f1}.  
A close-up of these features is presented in Figure~\ref{f4}.  
\emph{D-D$^\prime$} have a point-symmetric morphology, bending around 
\emph{IS} in arc-like features that extend from the \emph{IS} equator, 
at PA $\simeq$60$^{\circ}$ (240$^{\circ}$), to the \emph{IS} tips, at 
PA $\simeq$330$^{\circ}$ (150$^{\circ}$).  
These features do not look like filaments, but rather like broad 
($\simeq$ 1$''$), clumpy strips, particularly in the [N~{\sc ii}] 
image.  
The morphology is reminiscent of the low-excitation {\it polar caps} 
of NGC\,6543 \citep{MS92,B04} and of the {\it south fuzz} of NGC\,2392 
\citep{ODetal02}.

The surface brightness of these structures does not change significantly 
in the [N~{\sc ii}] line, but there is a clear enhancement of the H$\alpha$ 
and [O~{\sc iii}] emission in the equatorial regions that we denote as 
\emph{Dm-Dm$^\prime$}.  
To investigate these changes in more detail, we have used the well 
established [O~{\sc iii}]/H$\alpha$ vs.\ [N~{\sc ii}]/H$\alpha$ line 
diagnostic diagram.  
To construct this diagram, we have selected the rectangular regions 
encompassing \emph{D-D$^\prime$} and \emph{Dm-Dm$^\prime$} shown in 
Figure~\ref{f5}-{\it right} and computed the dereddened 
[O~{\sc iii}]/H$\alpha$ 
and [N~{\sc ii}]/H$\alpha$ ratios at every pixel with surface brightness 
above a threshold value of 3$\sigma$.  
The resulting values are plotted in the [O~{\sc iii}]/H$\alpha$ 
vs.\ [N~{\sc ii}]/H$\alpha$ diagrams shown in Fig.~\ref{f5}-{\it left}.  
In these diagrams, \emph{D} and \emph{D$^\prime$} present very similar 
distributions in their excitation conditions, and the same applies 
between \emph{Dm} and \emph{Dm$^\prime$}.  
These show slightly higher [O~{\sc iii}]/H$\alpha$ values than 
\emph{D-D$^\prime$}, but very different [N~{\sc ii}]/H$\alpha$ 
distributions, with \emph{Dm-Dm$^\prime$} values lower than 
$\sim$0.15, while \emph{D-D$^\prime$} present values of this 
ratio up to 0.6.  
Therefore, there is a notable segregation in the [O~{\sc iii}]/H$\alpha$ 
vs.\ [N~{\sc ii}]/H$\alpha$ plane between \emph{D-D$^\prime$} and 
\emph{Dm-Dm$^\prime$}, indicating different excitation conditions.

The different nature of \emph{D-D$^\prime$} and \emph{Dm-Dm$^\prime$} is 
further substantiated by their kinematics in the [N~{\sc ii}] echellogram 
at PA 60$^{\circ}$ (Fig.\ref{f3}).  
In this echellogram, four distinct compact knots are detected at the 
location of \emph{D-D$^\prime$} and \emph{Dm-Dm$^\prime$}.  
Of these four knots, the pair of knots located at the same side of 
the nebula have opposite velocities, i.e., \emph{D} and \emph{Dm} 
have very similar locations, but very different radial velocities, 
and the same applies to \emph{D$^\prime$} and \emph{Dm$^\prime$}.  
The position-velocity (PV) map of the [N~{\sc ii}]/H$\alpha$ ratio 
(Fig.~\ref{f3}) shows that the brighter two [N II] knots, corresponding to 
\emph{D-D$^\prime$}, have lower excitation, while the two fainter 
knots show higher excitation and can be identified with 
\emph{Dm-Dm$^\prime$}.  
Therefore, \emph{D} and \emph{Dm$^\prime$} are blueshifted, with systemic 
radial velocities of --21~km~s$^{-1}$ and --27~km~s$^{-1}$, respectively, 
while \emph{D$^\prime$} and \emph{Dm} are redshifted, with systemic radial 
velocities of +21~km~s$^{-1}$ and +27~km~s$^{-1}$, respectively.

The kinematics along \emph{D-D$^\prime$} has been further mapped at PAs 
150$^{\circ}$, 166$^{\circ}$ and 180$^{\circ}$.  
In the corresponding echellograms, \emph{D-D$^\prime$} appear as bright, 
compact knots with a velocity $FWHM$ of $\sim$18~km~s$^{-1}$ in the 
[N~{\sc ii}] $\lambda$6584 \AA\ line.  
\emph{D} is blueshifted and its radial velocity shows little variations 
along this structure as shown in Fig.~\ref{f4}.  
\emph{D$^\prime$} is redshifted and shows velocity variations symmetric 
to those presented by \emph{D}. 
It must be noted that these velocities are opposed to the line tilt 
shown by the inner shell, whose Northwest protrusion is redshifted, 
while its Southeast tip is blueshifted.

\begin{figure*}[!t]
\centerline{
~~~~~~~~~~~~~~~~~~\includegraphics[width=6.9cm,angle=0]{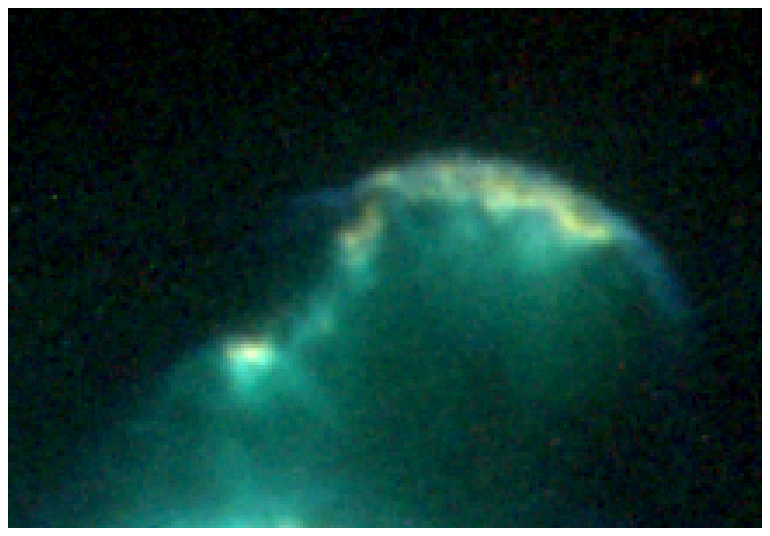}
~~~~~~~~~~~~~~~~\includegraphics[width=6.9cm,angle=0]{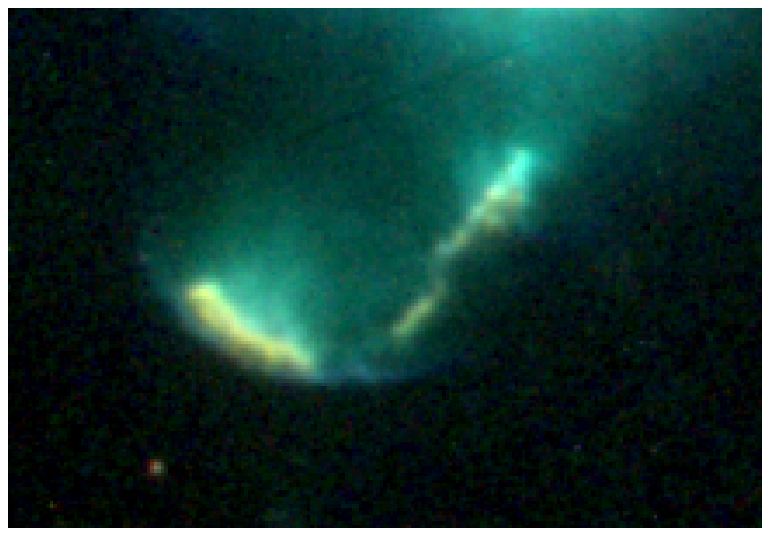}
}
\vspace*{0.2cm}
\centerline{
\includegraphics[width=17.5cm,angle=0]{f6c_IC4634.eps}
}
\caption{
\emph{HST} WFPC2 H$\alpha$ (green), [O~{\sc iii}] (blue), and [N~{\sc ii}] 
(red) composite color picture {\it (top)}, and [O~{\sc iii}]/H$\alpha$ 
{\it (center)} and [N~{\sc ii}]/H$\alpha$ {\it (bottom)} ratio maps of the 
bow shock features \emph{A} and \emph{B} {\it (left)} and \emph{A$^\prime$}
and \emph{B$^\prime$} {\it (right)}. 
The color pictures and ratio maps are displayed in logarithmic scale.  
As in Fig.~\ref{f1}, the greyscale of the ratio maps have lower limit of 0, and 
upper limits of 1.0 for the [N~{\sc ii}]/H$\alpha$ ratio maps and 4.0 for 
the [O~{\sc iii}]/H$\alpha$ ratio maps.  
}
\label{f6}
\end{figure*}

The physical conditions at the tips of \emph{D-D$^\prime$} have been 
derived using temperature and density sensitive line ratios obtained 
from the medium dispersion spectroscopy.  
Using the line intensities listed in Tab.\,2, we have computed an 
electron density of 3,800~cm$^{-3}$ for \emph{D} and 5,200~cm$^{-3}$ 
for \emph{D$^\prime$}, i.e., the density of \emph{D-D$^{\prime}$} is 
2--3 times lower than the density of the inner shell.  
The [N~{\sc ii}] electron temperature is found to be 11,500 K 
and 10,200~K for \emph{D} and \emph{D$^\prime$}, respectively. 
The [O~{\sc iii}] electron temperature is 10,500~K for \emph{D} 
and 10,000 K for \emph{D$^\prime$}.

\subsection{The System of Bow Shock-Like Structures}

The system of bow shock-like structures observed at PAs 
$\sim$150$^\circ$ (\emph{A-A$^\prime$}), 
$\sim$166$^\circ$ (\emph{B-B$^\prime$}), and 
$\sim$120$^\circ$ (\emph{C-C$^\prime$}) 
is one of the most remarkable features 
of IC\,4634.  
There are very few examples of bow shock structures among PNe 
\citep[e.g., IC\,4593,][]{Cetal96}.   
In the following subsections, we describe in further detail the structure, 
excitation, physical conditions, and kinematics of these features.

\subsubsection{Small-Scale Structure}

The H$\alpha$, [N~{\sc ii}] and [O~{\sc iii}] composite pictures of \emph{A} 
and \emph{A$^\prime$} shown in Figure~\ref{f6} reveal a wealth of small-scale 
structures in these features.  
The bow shock-like feature \emph{A} lies 10$^{\prime\prime}$ 
NW of the central star.  
The counter bow shock (feature \emph{A$^\prime$}) is precisely symmetric 
on the opposite side of the central star (see Fig.~\ref{f1}).  
A chain of knots extends towards the SE from structure \emph{A} and 
towards the NW from structure \emph{A$^\prime$}, ending in a smaller 
sized bow shock labeled \emph{B} and \emph{B$^\prime$} in the NW and 
SE, respectively, that lie at $\sim$7$^{\prime\prime}$ from IC\,4634 
central star.

Both in \emph{A} and \emph{A$^\prime$} there are well-defined bow-shaped 
structures with sharp edges that emit predominantly in [O~{\sc iii}] 
(shown in blue in Fig.~\ref{f6}-{\it top}).  
Fainter [O~{\sc iii}], [N~{\sc ii}] and H$\alpha$ diffuse 
emission is observed in the region between the bow shock 
and the inner nebula.  
In the brightest regions of \emph{A} and \emph{A$^\prime$}, several 
knots can be resolved appearing distinct in [N~{\sc ii}] (Fig.~\ref{f6}), 
but surrounded by diffuse emission in [O~{\sc iii}].  
The twisted distribution of these [N~{\sc ii}] knots is indicative 
of the development of instabilities at this location.

The bow shock structures \emph{B} and \emph{B$^\prime$} show faint 
wings that are also more prominent and extended in the [O~{\sc iii}] 
image.  
Only one knot, located at the cap of the bow shock structure, 
is resolved within \emph{B}.  
The [O~{\sc iii}] image shows that the spatial distribution of this 
knot is considerably more extended than the corresponding H$\alpha$ 
and [N~{\sc ii}] intensity peaks.  
In the emission line images of the counter bow shock \emph{B$^\prime$}, 
three knots are resolved.

\emph{C} and \emph{C$^\prime$} (Fig.~\ref{f1}) are a pair of low-emissivity 
features located along an axis at PA $\sim$120$^\circ$.  
In the [N~{\sc ii}] image, they show pointed, triangular tips, that 
are surrounded by round, cap-like structures in the [O~{\sc iii}] 
and H$\alpha$ images.  
At the outermost skin of these cap-like structures, the 
[O~{\sc iii}]/H$\alpha$ line ratio is enhanced.

\subsubsection{Excitation and Physical Conditions}

The [O~{\sc iii}]/H$\alpha$ and [N~{\sc ii}]/H$\alpha$ ratio maps show 
abrupt changes in these line ratios throughout \emph{A-A$^\prime$}, 
\emph{B-B$^\prime$}, and \emph{C-C$^\prime$} (Fig.~\ref{f6}).  
At the leading edge of the bow shock structures, the [O~{\sc iii}] 
emission is enhanced, with [O~{\sc iii}]/H$\alpha$ values of 2--3.  
At the brightest regions of \emph{A-A$^\prime$} and \emph{B-B$^\prime$} 
(i.e., at the emitting knots of the bow shock structure), the 
[O~{\sc iii}]/H$\alpha$ ratio declines to 1--1.5, while the [N~{\sc ii}] 
emission is enhanced, with [N~{\sc ii}]/H$\alpha$ values of 0.8--0.9, 
significantly raised from the value of 0.05--0.15 behind the bow shock.
Similarly, there are abrupt changes in these ratios across the 
\emph{B-B$^\prime$} features (Fig.~\ref{f6}), with the lowest 
[O~{\sc iii}]/H$\alpha$ ratios and the highest 
[N~{\sc ii}]/H$\alpha$ ratios at the location of the knots (i.e., 
at the head of the bow shock).  
The spatial offset between the [O~{\sc iii}] enhanced cap-shaped 
bow shocks and the low-ionization [N~{\sc ii}] bright knots is 
$\sim$0\farcs3.  
We note that the wings at the \emph{A-A$^\prime$} bow shocks show the 
highest excitation among these structures, with [O~{\sc iii}]/H$\alpha$ 
values of $\sim$4, and [N~{\sc ii}]/H$\alpha$ values lower than $\sim$0.2.

As in \S3.2, we have computed the [O~{\sc iii}]/H$\alpha$ vs.\ 
[N~{\sc ii}]/H$\alpha$ diagrams for the different rectangular 
regions encompassing \emph{A-A$^\prime$}, \emph{B-B$^\prime$}, 
and \emph{C-C$^\prime$} shown in Fig.~\ref{f5}-{\it right}.  
In these diagrams, shown in Fig.~\ref{f5}-{\it left}, it is noticeable the strong 
similarity between the distribution of the data points of each structure 
in the NW and SE regions.  
Data points from structures \emph{A-A$^\prime$} and \emph{B-B$^\prime$} 
tend to occupy an extended region in Fig.~\ref{f5}-{\it left}, implying large 
variations in the excitation conditions on small scales. 
Most points fall in the region of the diagram with [O~{\sc iii}]/H$\alpha$ 
values $\sim$2--3 and [N~{\sc ii}]/H$\alpha$ $\sim$0.05--0.20, corresponding 
to the gas behind the bow shock. 
There is a locus in this diagram characterized by low values of 
[O~{\sc iii}]/H$\alpha$ and large values of [N~{\sc ii}]/H$\alpha$, 
which correspond to the bright emitting knots. 
On the other hand, there are also points characterized by large values 
of the [O~{\sc iii}]/H$\alpha$ ratio, $\geq$3, and very low values of 
the [N~{\sc ii}]/H$\alpha$ ratio;  
these points correspond to the ''skin'' of [O~{\sc iii}] observed at the 
leading edge of the knots.  
The excitation of the bow shock-like structures \emph{B-B$^\prime$} is 
similar to that of \emph{A-A$^\prime$}.  
Most of the data points of \emph{B-B$^\prime$} fall in the region having 
[O~{\sc iii}]/H$\alpha$ in the range 2--3 and [N~{\sc ii}]/H$\alpha$ 
ranging from 0.05 to 0.20.  
The points with low [O~{\sc iii}]/H$\alpha$ values and high  
[N~{\sc ii}]/H$\alpha$ values corresponds to the [N~{\sc ii}] 
bright knots.  
Note the presence of several data points with large [O~{\sc iii}]/H$\alpha$ 
values (ranging from 3 to 4), which correspond to the leading edge of the 
bow shocks.

The data points from features  \emph{C-C$^\prime$} occupy a small 
region in the [O~{\sc iii}]/H$\alpha$ vs.\ [N~{\sc ii}]/H$\alpha$ 
diagram (Fig.~\ref{f5}-{\it right}). 
Most points fall in the region with large [O~{\sc iii}]/H$\alpha$ ratio 
($\sim$ 2.25 to 3.5) and low [N~{\sc ii}]/H$\alpha$ ratio ($\leq$ 0.15). 
Data points corresponding to the outer ``skin'' of [O~{\sc iii}] 
show [O~{\sc iii}]/H$\alpha$ larger than 3.5. 
There are noticeable differences between the distribution of the 
data points of \emph{A-A$^\prime$} and \emph{B-B$^\prime$} in the 
[O~{\sc iii}]/H$\alpha$ vs.\ [N~{\sc ii}]/H$\alpha$ diagram, and 
this of \emph{C-C$^\prime$}.  
In particular, the data points of \emph{C-C$^\prime$} have a smaller 
scatter in the [O~{\sc iii}]/H$\alpha$ vs.\ [N~{\sc ii}]/H$\alpha$ 
diagram than those of \emph{A-A$^\prime$} and \emph{B-B$^\prime$}, 
implying small variations in the excitation conditions across 
\emph{C-C$^\prime$}.  
Moreover, \emph{C-C$^\prime$} have higher excitation than 
\emph{A-A$^\prime$} and \emph{B-B$^\prime$}.

\begin{figure}[!t]
\centerline{
\includegraphics[width=9cm,angle=0]{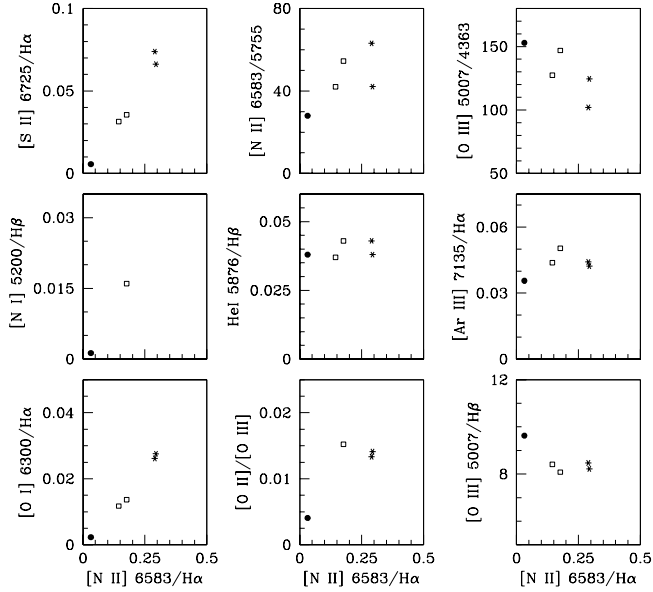}
}
\caption{
Different diagnostic diagrams for the inner shell \emph{IS} 
{\it (solid dots)}, the arc-like features \emph{D-D$^\prime$} 
{\it (open squares)}, and the bow shock-like features 
\emph{A-A$^\prime$} {\it (stars)} as derived from the 
undereddened line ratios measured in the low-dispersion spectra.  
}
\label{f7}
\end{figure}

Our ground-based medium-dispersion spectroscopy (Tab.~2) cannot resolve 
the rapidly changing ionization structure seen in the \emph{HST} images, 
but the use of additional line diagnostic diagrams from a variety of 
emission line ratios can provide further information on the excitation 
and physical conditions of \emph{A-A$^\prime$}.  
To better compare the overall excitation of \emph{A-A$^\prime$} 
with that of the inner shell, \emph{IS}, and \emph{D-D$^\prime$}, 
the line diagnostic diagrams shown in Figure~\ref{f7} also include 
emission line ratios from these regions.  
The spectra of \emph{A-A$^\prime$} show enhanced emission of 
low-excitation lines: 
the relative intensities of [O~{\sc i}], [O~{\sc ii}], [N~{\sc ii}], 
and [S~{\sc ii}] are up to 10 times higher than those in the spectrum 
of \emph{IS}, while the relative intensities of [O~{\sc i}], [N~{\sc ii}], 
and [S~{\sc ii}] are larger than these of \emph{D-D$^\prime$} by a factor 
of $\sim$2. 
The relative intensity of the [O~{\sc iii}] lines of both 
\emph{A-A$^\prime$} and \emph{D-D$^\prime$} is only 10$\%$--15$\%$ lower 
than these in \emph{IS}, while the He~{\sc i} and [Ar~{\sc iii}] lines 
have similar intensities among these different regions, indicating that 
the radiation field reaching both \emph{A-A$^\prime$} and \emph{D-D$^\prime$}, 
modified by the absorption in the IS, is similar in both cases and 
capable of ionizing species with ionization potential $\leq$ 30 eV.  
We have obtained the [O~{\sc iii}]/[O~{\sc ii}] intensity ratio, which 
is indicative of the ionizing parameter for a photoionized gas, 
for the three different regions.  
The largest value (i.e., the largest ionizing parameter) corresponds to 
the inner shell, as expected, with a value of $\sim$250.  
\emph{A-A$^\prime$} and \emph{D-D$^\prime$} show lower values in the 
narrow range from $\sim$70 to $\sim$80, thus indicating that these 
structures have similar local ionizing conditions (somewhat lower in 
\emph{A-A$^\prime$}).  
Even though the ionizing parameters are similar in \emph{D-D$^\prime$} and 
\emph{A-A$^\prime$}, the latter shows stronger emission of low excitation 
lines (see above), indicating shock excitation.

Fig.~\ref{f7} also reveals the progressive change in the [N~{\sc ii}] and 
[O~{\sc iii}] temperatures with distance from the central star of 
IC\,4634:  
the [N~{\sc ii}] 6584/5755 emission line ratio increases (and so the 
temperature inferred from this line ratio decreases) from \emph{IS} 
to \emph{A-A$^{\prime}$}, while the [O~{\sc iii}] 5007/4363 emission 
line ratio decreases (and so the temperature derived from this line 
ratio increases) as we move outwards from regions of low 
[N~{\sc ii}]/H$\alpha$ values at \emph{IS} to regions with larger 
values of this ratio.   
The electron temperatures of \emph{A} and \emph{A$^{\prime}$} are, 
respectively, 10,200~K and 12,200~K from the [N~{\sc ii}] lines, 
and 10,800~K and 11,400~K from the [O~{\sc iii}] lines.
Similarly, the electron density from the [S~{\sc ii}] doublet line ratio 
is found to be 2,800~cm$^{-3}$ in \emph{A} and 1,500~cm$^{-3}$ in 
\emph{A$^\prime$}, i.e., 3--7 times lower than in \emph{IS}.

\subsubsection{Kinematics}

\begin{figure*}[!t]
\centerline{
\includegraphics[width=17.5cm,angle=0]{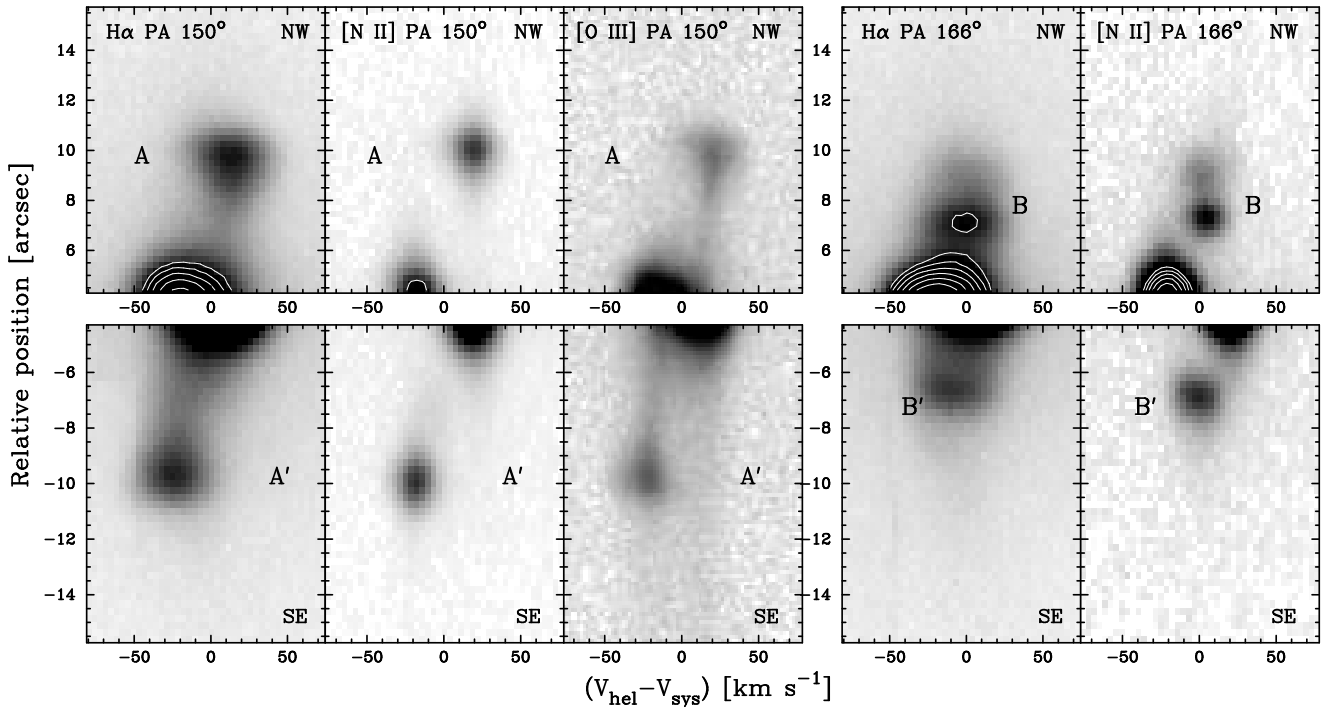}
}
\caption{
Echellograms of the H$\alpha$, [N~{\sc ii}] $\lambda$6584 \AA, and 
[O~{\sc iii}] $\lambda$5007 \AA\ emission lines along PA 150$^\circ$, 
and echellogram of the H$\alpha$ and [N~{\sc ii}] $\lambda$6584 \AA\ 
emission lines along PA 166$^\circ$ of the bow shock-like features 
\emph{A-A$^\prime$} and \emph{B-B$^\prime$}.  
}
\label{f8}
\end{figure*}

The large opening angle of these bow shock-like structures (\emph{A} and 
\emph{A$^\prime$} have almost flat morphologies) suggests a low inclination 
angle with the plane of the sky.
This is confirmed by the PV diagrams of the bow shock-like structures 
\emph{A-A$^\prime$} and \emph{B-B$^\prime$} displayed in Figure~\ref{f8} 
that show low radial velocities with respect to the systemic velocity: 
$\pm$20 km~s$^{-1}$ for \emph{A-A$^\prime$}, and $\pm$3 km~s$^{-1}$ 
for \emph{B-B$^\prime$}.  
The H$\alpha$ and [N~{\sc ii}] $\lambda$6584 \AA\ echellogram along 
PA=120$^{\circ}$ (not shown in Fig.~\ref{f8}) shows faint emission 
from \emph{C} and \emph{C$^\prime$} moving at radial velocities with 
respect to the systemic velocity of $\pm$30~km~s$^{-1}$.  
We note that \emph{A}, \emph{B}, and \emph{C} (\emph{A$^\prime$}, 
\emph{B$^\prime$}, and \emph{C$^\prime$}) are receding from 
(approaching) us, opposite to the motion of \emph{D} 
(\emph{D$^\prime$}).

It is also interesting to note the detailed line shape in the PV 
diagrams of Fig.~\ref{f8} at \emph{A-A$^\prime$} and \emph{B-B$^\prime$}.  
At the location of \emph{A-A$^\prime$}, the H$\alpha$ and [O~{\sc iii}] 
emission lines are wedge-shaped with the widest side at the line tip 
(the leading edge of the bow shock), while the [N~{\sc ii}] line shape 
is nearly round.  
The FWHM of the H$\alpha$ emission line of both \emph{A-A$^\prime$} 
increases from $\sim$30 km~s$^{-1}$ at the narrow end of this 
wedge-shaped feature to $\sim$70 km~s$^{-1}$ at its wide end.  
A close examination of the H$\alpha$ and [O~{\sc iii}] 5007 \AA\ 
PV diagrams reveals a velocity gradient at the leading edge of 
\emph{A-A$^\prime$}, whereas the velocities in the region behind 
the bow shock seem to be rather constant.   
The lowest velocities are found at the leading edge of the 
bow shock, facing away from the central star of IC\,4634, 
where the radial velocity of the H$\alpha$ line decreases 
$\sim$10~km s$^{-1}$ within 1$^{\prime\prime}$.
Across \emph{B-B$^\prime$}, the radial velocity is roughly constant.  
The H$\alpha$ emission line profiles show a mean FWHM of 
$\sim$40~km s$^{-1}$. 
The emission ends in a sharp drop at the outer edge of \emph{B-B$^\prime$}.

Finally, the string of [N~{\sc ii}] knots that arises from the edge of the 
\emph{A} and \emph{A$^\prime$} bow shocks may be suggestive of a trail of 
material left in their motion forward.  
Indeed, the radial velocity in the outer low-ionization arcs shows an 
abrupt change from the bow shock structure to the linear string of knots, 
suggesting a deceleration.  
We will show in $\S4.2$ that the interaction between a precessing 
collimated outflow with time-dependent velocity and the surrounding 
medium offers an alternative explanation to the string of [N~{\sc ii}] 
knots.

\subsection{The [O~{\sc iii}] Skin}

It has been mentioned that the [O~{\sc iii}]/H$\alpha$ ratio is 
clearly enhanced in cap-like structures just outside the bow shocks 
\emph{A-A$^\prime$}, \emph{B-B$^\prime$}, and \emph{C-C$^\prime$}.  
The [O~{\sc iii}]/H$\alpha$ ratio map (Fig.~\ref{f1}-{\it right-center}) 
reveals that these [O~{\sc iii}] enhanced cap-like structures extend 
on a strip that inscribe the inner regions of IC\,4634, like a thin 
shell or skin of enhanced [O~{\sc iii}] emission.  
Only the outer envelope (\emph{Env}) and helical structures 
(\emph{H} and \emph{Hd}) are located outside this skin of enhanced 
[O~{\sc iii}]/H$\alpha$ ratio.

The presence of a skin of enhanced [O~{\sc iii}]/H$\alpha$ ratio 
is rare among PNe;  only in NGC\,6543 a structure of this kind has 
been reported \citep{B04}.  
In his detailed analysis of narrow-band \emph{HST} WFPC2 images of 
NGC\,6543, Balick noticed that there is no observational artifact 
(contamination of [N~{\sc ii}] emission in the H$\alpha$ image or 
incorrect correction for wavelength-dependent geometric distortions 
in the camera optics), nor ionization effects or local variations 
of the O/H abundances or the electronic temperature in NGC\,6543 
able to produce the observed increase of the [O~{\sc iii}]/H$\alpha$ 
ratio at the nebular edge.  
Observational artifacts can also be dismissed as the origin of the 
observed morphology in the [O~{\sc iii}]/H$\alpha$ ratio map of 
IC\,4634, as the emission in [N~{\sc ii}] is much weaker than in 
H$\alpha$, as well as a local variation of the O/H abundances which 
are rather constant throughout the nebula.  
Emission from the main nebula scattered in the PN outer layers can 
also be excluded as the [O~{\sc iii}]/H$\alpha$ ratio in this skin 
is significantly different from the ratio value measured in the 
bright, innermost regions of the nebula.  
The only plausible cause of the observed enhancement in 
the [O~{\sc iii}]/H$\alpha$ ratio is a local increase of 
O$^{++}$/H$^+$ or $T_{\rm e}$.

The origin of the skin of enhanced [O~{\sc iii}]/H$\alpha$ ratio 
in NGC\,6543 was concluded to be uncertain \citep{B04}.  
In IC\,4634, this structure is related to the bow shock 
structures \emph{A-A$^\prime$}, \emph{B-B$^\prime$}, and 
\emph{C-C$^\prime$} and, therefore, it can be speculated that 
it originates in the interaction of fast collimated outflows 
with the nebular material.  
The shocks produced by fast collimated outflows would excavate a cavity 
in the low density nebular envelope and propagate outwards, inducing a 
marginal increase of the electronic temperature in a forward shock that 
enhances the [O~{\sc iii}] as also observed in wind-blown bubbles around 
WR stars \citep{Getal00}.  
A similar origin can be attributed to such structure in NGC\,6543, as 
suggested by the caps of enhanced [O~{\sc iii}]/H$\alpha$ associated 
with the jet-like features of NGC\,6543 (see Figure~2 of Balick 2004).

\subsection{The Outer Envelope}

The \emph{HST} images of IC\,4634 reveal an envelope of faint emission 
surrounding its central regions that we have labeled as \emph{Env} in 
Fig.~\ref{f1}.  
This envelope has a patchy appearance, with a distinct arc towards the 
East, and individual filamentary (\emph{F-F$^\prime$}) and arc-like 
(\emph{G}) features towards the West.  
Some of these features are best seen in the [O~{\sc iii}]/H$\alpha$ 
ratio map shown in Fig.~\ref{f1}-{\it right-bottom}.  
Overall, the morphology of the envelope can be classified as elliptical, 
with its major axis along PA$\sim$120$^\circ$.

Since the envelope is more clearly detected in the H$\alpha$ and 
[O~{\sc iii}] images than in the [N~{\sc ii}] image, it suggests 
a high excitation.  
Its surface brightness is low, up to $\sim$1,000 lower than 
that of the inner shell in the H$\alpha$ line.  
If we assume the same electron temperature as in the inner shell, 
its density can be scaled down from that of the inner shell to 
$\approx$100~cm$^{-3}$.

Finally, the envelope is detected in the high-dispersion spectra, 
especially in the echellograms of the H$\alpha$ line.  
Its emission shows a broad, unresolved line with a radial velocity 
similar to the systemic velocity.  
This structure seems to be an inert irregularly shaped, low-density 
outer envelope.

\subsection{The Outer Helical Structure}

\begin{figure}[!t]
\centerline{
\includegraphics[width=8.5cm,angle=0]{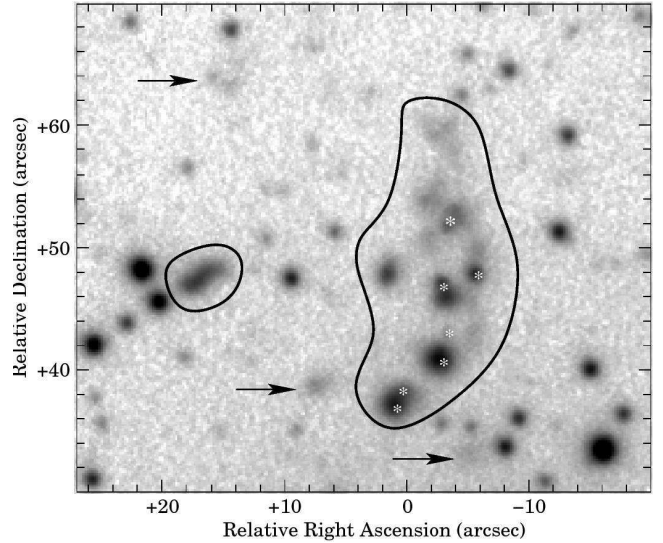}
}
\caption{
Grey scale representation of the NOT H$\alpha$+[N~{\sc ii}] image of 
the string of faint knots forming component \emph{H} (see Fig.~\ref{f1}).  
The arrows and closed curves overlaid on the image mark the 
nebulosities associated with this component, while the white 
stars indicate the position of field stars overimposed onto 
some of the line emission knots, as identified by comparing 
this narrow-band image with an \emph{HST} WFPC2 broad-band 
$V$ image.  
The origin (0,0) is at the position of the central star of IC\,4634.
}
\label{f9}
\end{figure}

The existence of a distant string of faint knots associated with 
IC\,4634 was first reported by \citet{GMC04}.  
This feature, labeled \emph{H} and \emph{Hd} in Fig.~\ref{f1}-{\it right-top}, 
is shown in greater detail in the deep NOT H$\alpha$+[N~{\sc ii}] image 
presented in Figure~\ref{f9}.  
In this figure, \emph{H} and \emph{Hd} are composed of several knots 
detected from PA $\simeq$$-$5$^{\circ}$ to 20$^{\circ}$ at angular 
distances $37''$--$64''$ from the central star of IC\,4634.  
In the \emph{HST} images, the knots in the structure \emph{H} 
are resolved into compact cores surrounded by clumpy emission 
or show the appearance of a string of knots.  
Morphologically, component \emph{H} forms an arc-like string of knots 
whose orientation is different from that of the major nebular axis, but 
similar to this of components \emph{B-B$^\prime$} and \emph{D-D$^\prime$}.  
Component \emph{Hd} consists of several faint diffuse knots, 
including a large ($\sim$ 25$''$) corkscrew-shaped structure 
oriented along the South-North direction.  
We have obtained deep narrow-band images to search for a possible 
Southern counterpart of \emph{H} and \emph{Hd}, but our search has 
yielded negative results.

The H$\alpha$ and [N~{\sc ii}] emission from \emph{H} and \emph{Hd} 
is detected in the high-dispersion spectroscopic observations.  
The emission lines can be fit with a single Gaussian profile with 
$FWHM$ of $\sim$25 km~s$^{-1}$ in the H$\alpha$ line and $\sim$20 
km~s$^{-1}$ in the [N~{\sc ii}] $\lambda$6584 line.  
The systemic radial velocity derived from these lines is 
very similar to the nebula systemic radial velocity.

\section{A Precessing Collimated Outflow in IC\,4634}

\subsection{Observational Evidence and Similarities with Herbig-Haro Objects}

The morphology of the bow-shaped features \emph{A-A$^\prime$}, 
\emph{B-B$^\prime$}, and \emph{C-C$^\prime$} resembles that of  
Herbig-Haro (HH) jets.  
These jets are morphologically characterized by chains of aligned knots with 
bow shock-like appearance, and a leading bow shock, the ``head'' of the 
jet, where the supersonic flow slams into the surrounding material.  
Observations of the knots of HH jets have been interpreted by several 
authors as the result of time-dependent variations in the velocity of 
the flow \citep[see, e.g.,][]{Retal90}.  
Such a jet model with variable ejection velocity has been shown 
to reproduce the overall morphology and kinematical properties of 
the high-velocity jets of the proto-planetary nebula Hen\,3-1475 
\citep{VRR04}.  
The bow shock structure of \emph{A-A$^\prime$}, \emph{B-B$^\prime$}, 
and \emph{C-C$^\prime$} is consistent with those features being the 
result of a jet, ejected with variable velocity, interacting with the AGB 
remnant.

The suggestion drawn from the morphology of IC\,4634 is reinforced by 
the kinematics of \emph{A-A$^\prime$}. 
Its H$\alpha$ and [O~{\sc iii}] PV diagrams clearly show a 
wedge-shaped feature analogous to those observed in several 
HH objects \citep[e.g.,][]{BS85}.  
A wedge-shaped feature in the PV diagram is expected for a bow shock 
moving nearly on the plane of the sky, as predicted by the traditional 
``3/2''-dimension bow shock model which is a reasonable approximation 
to the leading bow shock of an HH jet \citep{RB85,RB86,HRH87}.  
In such case, the radial velocity dispersion increases suddenly at 
the stagnation point of the bow shock, producing a line broadening 
that looks like an horizontal edge in the PV diagram.  
The amount of broadening of the emission line profiles can be related to 
the velocity of the bow shock \citep{HRH87}.  
In the case of \emph{A-A$^\prime$}, the observed width of the line 
would indicate a bow shock velocity of $\sim$ 100 km s$^{-1}$. 
We conclude that the PV diagrams of \emph{A-A$^\prime$} are, at least 
qualitatively, compatible with the predictions of a leading bow shock 
(i.e., the ``head'' of a jet) nearly on the plane of the sky.

Further support for this scenario is provided by the spectral properties 
of \emph{A-A$^\prime$}.  
The dereddened emission line ratios in these regions with intense 
emission in low-ionization lines are reminiscent of shock-excited 
nebula such as HH objects.  
While shocks might be collisionally exciting the low ionization 
emission lines of \emph{A-A$^\prime$}, a detailed comparison of 
the observed emission line ratios with spectra predicted by 
photoionized shock models \citep[e.g.,][]{D97} is not 
straightforward, since the bow shocks in IC\,4634 are illuminated 
by the stellar ionizing flux from the post-shock direction.  
A more realistic comparison is enabled by the axisymmetric numerical 
simulations of a shocked, dense cloudlet moving away from a source 
of ionizing photons through a uniform and photoionized environment 
that have been recently obtained by \citet{RR07} and \citet{Retal07}.  
These simulations not only produce synthetic spectra that are 
qualitatively consistent with the spectra of \emph{A-A$^\prime$}, 
but also reproduce the ionization stratification of 
\emph{A-A$^\prime$}, with the gas in the downstream region being 
more highly ionized than the outward-facing edge and the 
[O~{\sc iii}] emission showing a larger extension towards the 
photoionizing source.

It is thus tempting to interprept the point-symmetry morphology of 
IC\,4634 as the result of moderate velocity jets with a time-dependent 
ejection velocity.  
Furthermore, the point-symmetric morphology of IC\,4634, which gives 
its S-shaped appearance to the nebula, and the varying sign of the 
radial velocity are commonly interpreted as the direct result of a 
time-dependent direction of ejection of the source.    
In the next section, we explore into further detail this supposition that 
can, in general, be of interest to other point-symmetric PNe 
\citep{Cetal95,LS03,GF04,VRR04,SB06}.

\subsection{Modeling a Precessing Collimated Outflow in IC\,4634}

\subsubsection{Initial Conditions and Numerical Methods}

The morphology and kinematics derived for IC\,4634 place important 
constrains and clues on some of the parameters describing the motion 
of such a precessing outflow.  
The bow shock morphology of \emph{A-A$^\prime$}, 
and the spatial distribution of the [O~{\sc iii}] and H$\alpha$ 
emission, preceding the [N~{\sc ii}] emission, reveals shock 
excitation.  
The observed radial velocities are low, suggesting that the angle between 
the precession axis and the plane of the sky, the inclination angle $i$, 
is small.  
We will thus assume that the precession axis is on the plane of the sky.  
Finally, the semi-angle of the aperture of the precession cone, $\alpha$, 
can be derived by measuring the angle subtended by \emph{A} and \emph{B} 
with respect to the central star.  
This angle is estimated to be $\sim$11\arcdeg.

Numerical simulations were carried out with the 3D code Yguaz\'u-a 
\citep[][]{RNV00,Retal02}, using a 5-level binary adaptive grid.
The dimensions of the computational domain are 2.2$\times$10$^{17}$~cm, in 
the x- and y-directions, and 4.4$\times$10$^{17}$~cm in the z-direction, 
i.e., 6\farcs3$\times$12\farcs7 at a distance of 2.3 kpc, with a maximum 
resolution of 8.6$\times$10$^{14}$~cm, i.e., $\sim$0\farcs025.  
This code integrates the gas-dynamic equations by using the ''flux vector 
splitting'' scheme of \citet{VL82}.  
Together with the gas-dynamic equations, several rate equations for the
atomic/ionic species are also integrated.  
These species are: 
H~{\sc i}, H~{\sc ii}, He~{\sc i}, He~{\sc ii}, He~{\sc iii}, C~{\sc ii}, 
C~{\sc iii}, C~{\sc iv}, N~{\sc i}, N~{\sc ii}, N~{\sc iii}, O~{\sc i}, 
O~{\sc ii}, O~{\sc iii}, O~{\sc iv}, S~{\sc ii}, and S~{\sc iii} 
\citep[see details about the reaction and the cooling rates in][]{Retal02}.  
These rate equations enable the computation of a non-equilibrium 
function for the radiative losses.
Given a set of initial conditions for the jet and surrounding 
circumstellar medium (CSM), Yguaz\'u-a determines the temperature 
and density distributions at a given time.  
The temperature and density distributions allow us to compute the 
emission line coefficients of the H$\alpha$, [N~{\sc ii}] $\lambda$6584, 
and [O~{\sc iii}] $\lambda$5007 emission lines.  
The intensity of the H$\alpha$ line is computed considering 
the contributions from the recombination cascade and from 
$n=1\to3$ collisional excitations. 
The intensity of the forbidden lines [N~{\sc ii}] $\lambda$6584 and 
[O~{\sc iii}] $\lambda$5007 are calculated by solving 5-level atom 
problems, using the parameters of \citet{M83}.  
These intensities can be integrated along the line of sight 
to produce synthetic emission maps and PV diagrams for a 
wide slit covering completely the working surfaces.

A jet of number density $n_{\rm j}$ is injected in the base of the 
computational domain ($z=0$) at position (1.1$\times$10$^{17}$~cm, 
1.1$\times$10$^{17}$~cm) in the $xy$-plane, with a radius and length 
of 2.6$\times$10$^{15}$~cm and 4.3$\times$10$^{15}$~cm, respectively.  
The jet density is assumed to be 10$^4$ cm$^{-3}$.  
The jet velocity is modeled by the relationship:
\begin{equation}
v_\mathrm{j} = v_0 + \Delta v\ \sin \bigg(2 \pi \frac{t}{\tau_\mathrm{v}}\bigg)
\label{vjet}
\end{equation}
\noindent 
where $v_0$ is the mean velocity, $\Delta v$ is the amplitude 
of the velocity variation, $t$ is the time, and $\tau_{\rm v}$ 
is the period of the velocity variation.  
The symmetry axis of this jet precesses with period $\tau_{\rm p}$.

The surrounding CSM has been assumed to be produced by an AGB wind of 
constant mass-loss rate, $\dot M$, and expansion velocity, $v_{\rm AGB}$.  
The dependence of the density of such medium with the radial distance 
to the central star, $r$, is given by: 
\begin{equation}
\rho(r) = \rho_0 \bigg(\frac{r_0}{r}\bigg)^2 
\label{rhocsm}
\end{equation}
where the density, $\rho_0$, at radius $r_0$ is determined by the 
mass loss-rate and expansion velocity of the AGB wind as 
$\rho_0=\dot{M}/(4\,{\pi}\,r_0^2\,v_{\rm AGB})$.  
In our simulations, the mass-loss rate and expansion velocity of the 
AGB wind were assumed to be $10^{-6}$ $M_\odot$~yr$^{-1}$ and 20 
km~s$^{-1}$, respectively.  
The temperature of the CSM has been set to 100~K.

\subsubsection{Results of the Numerical Simulations}

\begin{figure}
\centerline{
\includegraphics[height=11cm,angle=0]{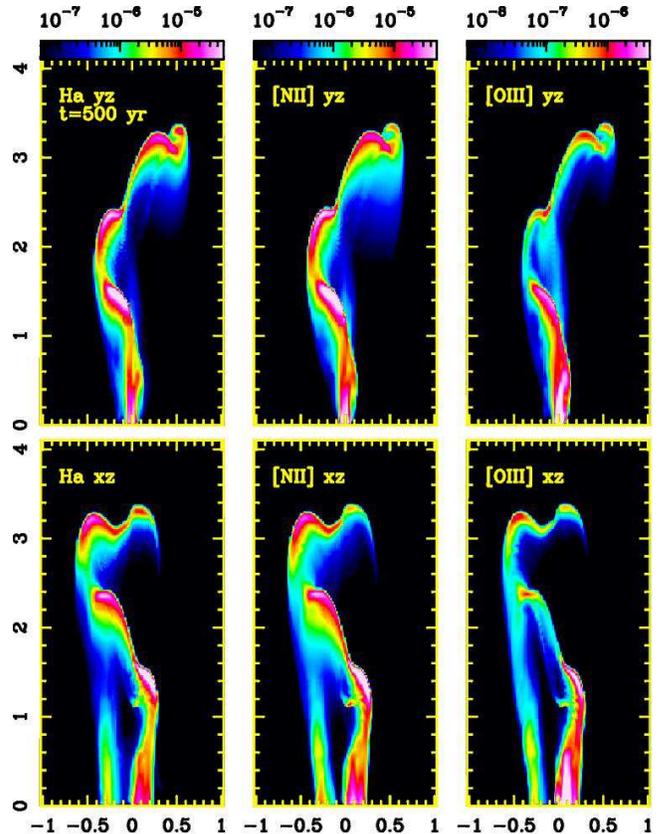}
}
\caption{
H$\alpha$ {\it (left)}, [N~{\sc ii}] {\it (center)}, and [O~{\sc iii}] 
{\it (right)} simulated emission maps of a velocity time-dependent 
precessing collimated outflow at a time of 500 yr. 
The uppers panels show the simulated emission maps for the 
$yz$--projection, while the $xz$--projection is displayed 
in the bottom panels.  
The spatial axes show the distance to the central star of IC\,4634 
in units of $10^{17}$~cm.
The maps are displayed in logarithmic scale with the flux level in 
units of ergs~s$^{-1}$~cm$^{-2}$~str$^{-1}$ shown in the colorbar 
located at the top of each upper panel.  
}
\label{f10}
\end{figure}

\begin{figure}
\centerline{
\includegraphics[height=11cm,angle=0]{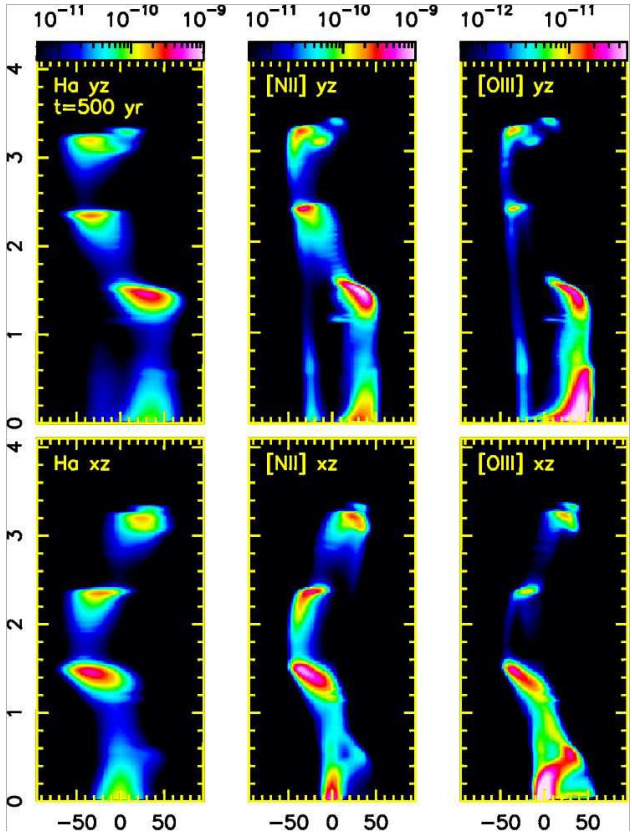}
}
\caption{
Simulated PV diagrams for the H$\alpha$ {\it (left)}, [N~{\sc ii}] 
{\it (center)}, and [O~{\sc iii}] {\it (right)} emission lines of 
a velocity time-dependent precessing collimated outflow at a time 
of 500 yr.  
The bottom panels display the $xz$--projection, while the upper panels 
show the $yz$--projection.  
In the panels, the horizontal axis displays the systemic radial 
velocity in units of \kms, and the vertical axis the distance 
to the central star of IC\,4634 in units of $10^{17}$~cm. 
The maps are displayed in logarithmic scale with the flux level in 
units of ergs~s$^{-1}$~cm$^{-2}$~str$^{-1}$~(cm~s$^{-1}$)$^{-1}$ 
shown in the colorbar located at the top of each upper panel.  
}
\label{f11}
\end{figure}

Several numerical simulations were carried out varying $v_{\rm j}$, 
$\Delta v$, $\tau_{\rm p}$, and $\tau_{\rm v}$, to produce synthetic 
emission maps and PV diagrams that can be directly compared to the 
observed images and long-slit echellograms of \emph{A-A$^\prime$}, 
\emph{B-B$^\prime$}, and \emph{D-D$^\prime$}.  
There is a qualitative agreement between the synthetic and 
observed images and PV diagrams for $v_0$=300~km~s$^{-1}$, 
$\Delta v$=25~km~s$^{-1}$, $\tau_{\rm p}$=460~yr, and $\tau_{\rm v}$=110 
yr (Figures~\ref{f10} and \ref{f11}).  
The jet expansion velocity implies a jet mass-loss rate of 
$\sim$2$\times$10$^{-7} M_\odot$~yr$^{-1}$.

Fig.~\ref{f10} displays the H$\alpha$, [N~{\sc ii}], and [O~{\sc iii}] 
simulated emission maps obtained for a calculation time of 500 yr. 
At this time, the jet structure achieves a length of 
3.3$\times$10$^{17}$~cm, which corresponds to an angular 
size of 9\farcs8 at a distance to IC\,4634 of 2.3 kpc.  
The upper panels correspond to the $yz$-projection (i.e., the line 
of sight is along the $x$-axis), while the bottom panels represent 
the $xz$-projection (i.e., the line of sight is along the $y$-axis), 
basically meaning the initial direction along which the jet is 
emitted.  
Several working surfaces are observed in these maps at 
distances of 1.4$\times$10$^{17}$~cm, 2.4$\times$10$^{17}$~cm, 
and 3.3$\times$10$^{17}$~cm, which are produced by the 
variability of the jet velocity, when slow gas is swept up by 
fast material.  
The observed morphology and spatial distribution of \emph{A-A$^\prime$} 
and \emph{B-B$^\prime$} are more closely reproduced by the $yz$-projection 
than by the $xz$-projection.  
We notice that the spatial separation seen at the head of the 
\emph{A-A$^\prime$} features between the H$\alpha$ and 
[O~{\sc iii}] emissions are not reproduced by the synthetic 
emission maps because the spatial structure of the region 
behind the shock, where the [O~{\sc iii}] emission arises 
mainly, is poorly resolved in our simulations.

Fig.~\ref{f11} displays the H$\alpha$, [N~{\sc ii}], and [O~{\sc iii}] 
simulated PV diagrams obtained at the same calculation time as 
Fig.~\ref{f10}.  
The centroids of the working surfaces in these PV diagrams do not 
show exactly the alternate positive and negative velocities observed 
in \emph{A-A$^\prime$} and \emph{B-B$^\prime$}, although they display 
similar sinusoidal variations with respect to the systemic velocity. 
This indicates that the initial direction along which the 
jet is emitted is not coincident with neither the $yz$- 
and $xz$-projections, or that the inclination angle differs 
from the one that has been assumed.  
On the other hand, a close examination of the PV diagrams in 
Fig.~\ref{f11} reveals a good agreement with the detailed kinematics 
of the working surfaces:  
the velocity widths or $FWHM$ of the synthetic emission lines are 
comparable to the observed ones, with larger values of the $FWHM$ 
for the H$\alpha$ line, and the wedge-shaped tips of the H$\alpha$ 
and [O~{\sc iii}] emission line profiles of \emph{A-A$^\prime$} are 
also well reproduced.

\section{Discussion}

\subsection{Multiple Shaping Agents in IC\,4634}

The data presented in the previous sections reaffirm the known 
complexity of IC\,4634.  
The nebula is composed of a series of morphological components 
that clearly reveal the action of different shaping agents.  
Among the prevalent shaping mechanisms in IC\,4634, we shall consider 
the interaction of the current fast stellar wind with the slow AGB wind, 
the ionizing flux of photons from its central star, and a series of fast 
collimated outflows.  
We discuss below the effects that the fast stellar wind and fast 
collimated outflows have had in the shaping of IC\,4634.

\emph{IUE} observations of the central star of IC\,4634 uncovered prominent 
P~Cygni profiles in the C~{\sc iii}, C~{\sc iv}, N~{\sc v}, and O~{\sc v} 
lines superposed on the stellar continuum.  
These P~Cygni profiles manifest the presence of a fast stellar wind whose 
terminal velocity has been derived to be 3,500--4,000 km~s$^{-1}$ 
\citep{HAF99}.  
This fast stellar wind has excavated a central cavity within the 
nebula, compressing the nebular material into a thin shell.  
This thin shell and the central cavity can be identified with the 
inner shell \emph{IS}, while the nebular envelope can be associated 
with the faint shell \emph{OS} surrounding this inner shell.  
In this respect, the double-shell morphology of the inner regions 
of IC\,4634 compares well with other PNe (e.g., NGC\,6826).

The bow shock morphology and spatial distribution of \emph{A-A$^\prime$}, 
\emph{B-B$^\prime$}, and \emph{C-C$^\prime$} are highly indicative of the 
interaction of precessing collimated outflows with surrounding material.  
Indeed, the detailed morphology and kinematics of \emph{A-A$^\prime$} 
and \emph{B-B$^\prime$} are successfully reproduced assuming the 
interaction of a precessing jet with a time-dependent velocity 
in the range of 300~km~s$^{-1}$ and a precession period $\sim$460 
yrs.  
This jet is interacting with material in the outer regions of 
IC\,4634 that forms the envelope, \emph{Env}, a region of rough 
elliptical symmetry that may represent an episode of major mass 
loss prior to the one that formed IC\,4634 inner regions.  
In their interaction, the collimated outflows generate bow-shocks 
structures and drive a forward shock that produces a high excitation 
skin.

One of the new nebular components revealed in this work is the 
outermost features \emph{H} and \emph{Hd}.  
The location of \emph{H} and \emph{Hd} with respect to IC\,4634 and 
their similar radial velocities make it unlikely that these components 
are unrelated to the nebula.   
The morphology of the \emph{H} feature, composed of a series of knots 
distributed along an arc-like structure, can be interpreted as a loop 
on the surface of an imaginary cone whose vertex is coincident with 
the central star of IC\,4634.  
Therefore, \emph{H} and \emph{Hd} may correspond to an ancient 
precessing ejection.  
It is worth emphasizing that the precession axis of this ancient 
ejection would be almost orthogonal to the axis of the precessing 
ejection that has been considered for \emph{A-A$^\prime$} and 
\emph{B-B$^\prime$}.

There are, thus, clear signs that IC\,4634 innermost shell, \emph{IS}, 
and the envelope, \emph{Env}, formed by the action of the current fast 
stellar wind on material associated to episodes of mildly asymmetric 
mass loss produced during the late AGB, while the bow shock structures 
\emph{A-A$^\prime$}, \emph{B-B$^\prime$}, and \emph{C-C$^\prime$}, and 
the arc-like features \emph{H} and \emph{Hd} have been produced by the 
action of fast, precessing collimated outflows.  
These collimated outflows may have also played some role in the shaping 
of the innermost regions of IC\,4634, although the evidence is not 
overwhelming.  
The inner shell of IC\,4634, \emph{IS}, is elongated and shows morphological 
and kinematical signs of blowout along its tips.  
While the origin of the asymmetry of \emph{IS} can be attributed 
to a density enhancement in the equator of the nebular envelope 
as suggested by the bright emission in the surrounding material 
along the minor axis in the H$\alpha$ and [O~{\sc iii}] images, 
it is worthwhile noting that the orientation and inclination of 
the collimated outflows associated with \emph{A-A$^\prime$} are 
coincident with those of the blowout along the tips of the inner 
shell.  
Therefore, it is possible that the collimated outflows 
responsible of \emph{A-A$^\prime$} may have been involved in 
the emergence of the blowout structures of \emph{IS} and in the 
development of the axisymmetry of this shell.  
This may also be the case of the inner shells of several PNe with 
collimated outflows (e.g., NGC\,6210 and NGC\,6884) that also show 
the mild point-symmetry exhibited by the inner shell of IC\,4634.

Similarly, the origin of the \emph{D-D$^\prime$} features may be 
associated with the action of collimated outflows.  
The kinematics and morphology of \emph{D-D$^\prime$} are very 
similar to those of the low-ionization {\it polar caps} described 
in NGC\,6543 \citep{MS92,B04}, although these structures may have 
different inclinations. 
A thorough discussion of the possible origins 
of knots and low-ionization structures in PNe is 
provided by \citet{ODetal02}.  
As in NGC\,6543, the \emph{D-D$^\prime$} arcs of IC\,4634 can be 
interpreted as intrusions of low-ionization, high-density knots 
that are expanding with the shell that surrounds its inner shell 
\citep{BH04}.  
These features may represent the relic of the interaction 
of a collimated outflow with the nebular envelope.  
On the other hand, \emph{Dm-Dm$^\prime$}, that has higher excitation 
than \emph{D-D$^\prime$}, and that are separated from these in the 
velocity space, can be related to the density enhancement around the 
equatorial region of \emph{IS}.

\subsection{Bow Shocks in Planetary Nebulae}

PNe display a large variety of low-ionization structures of which 
fast collimated outflows deserve special attention because of their 
outstanding kinematical properties \citep[e.g.,][]{GCM01}.  
The origin and nature of the fast collimated outflows seen in PNe 
has been disputed, but the morphological, kinematical, and physical 
properties of some of them seem to be suggestive of high-density 
bullets moving supersonically through the nebular material 
\citep{Betal94}.  
If this were the case, the interaction of the PN fast collimated outflows 
with the nebular material should lead to the formation of bow shocks whose 
morphologies would share some similarities with these of HH objects.  
Bow shocks would then be expected to be common among PNe with fast 
collimated outflows; 
the number of bow shocks identified in PNe, however, is small.  
In IC\,4634, we find one of the rare cases of bow shocks in PNe.  
The detailed morphology and kinematics of the bow shock structures 
\emph{A-A$^\prime$} and \emph{B-B$^\prime$} of IC\,4634 can be 
unambiguously ascribed to a precessing fast collimated outflow, 
and the morphology and ionization of the bow shock structures 
\emph{C-C$^\prime$} are reminiscent of the interaction of a fast, 
dense bullet with the nebular material.  
In view of these findings, it is worthwhile to revisit other cases 
of bow shocks in PNe.

So far, the best studied case of a bow shock in a PN is that of IC\,4593 
\citep{Cetal96}.  
In this nebula, we find several systems of knots and inward facing tails 
that have been interpreted as multiple collimated outflows propagating 
along different directions.  
These knots are preceded by outward facing caps with bow shock 
morphologies.  
The bow shock structures in IC\,4593 share many similarities with those 
of IC\,4634.  
Both have low radial velocities, show a decrease in the observed radial 
velocity and the broadening of the emission profile, and display caps 
of enhanced [O~{\sc iii}] emission as expected in bow shock models.  
All these properties suggest the motion of medium velocity jets near the 
plane of the sky that are interacting with the surrounding material.

The \emph{ansae} of NGC\,7009 are another promising example of fast, 
dense bullets ramming through the outer shells of the nebula.  
This interpretation was questioned as the ionization state of 
the gas declines at the head of the bow shock \citep{Betal98}, 
while the opposite is expected in bow shocks \citep{HRH87}. 
However, the observed ionization gradient can be reproduced by the 
numerical simulations of a shocked cloudlet moving away from the 
central star if the stellar ionizing photon flux, which modifies 
the ionization and excitation structure of the shock, is included 
\citep{RR07,Retal07}.  
In their study of IC\,4593 bow shocks, \citet{Cetal96} noted that 
the ionizing flux from the central star can play a predominant 
role in the ionization stratification of bow shocks in PNe.  
At any rate, an examination of a deep \emph{HST} WFPC2 
[O~{\sc iii}]/H$\alpha$ ratio map of NGC\,7009 reveals that 
the \emph{ansae} of NGC\,7009 are preceded by caps of 
enhanced [O~{\sc iii}] as expected in bow shocks 
\citep{Metal07,Getal08}.

Besides IC\,4593, IC\,4634, and NGC\,7009, there are very few other 
cases of bow shock associated with PNe.  
The [O~{\sc iii}]/H$\alpha$ ratio image of NGC\,6543 \citep{B04} shows 
a bow shock structure enveloping the precessing-like collimated outflows 
along the polar directions of this nebula \citep{MS92}, although this 
structure is not discussed in detail.  
A bow shock structure has also been described in NGC\,6572, although 
the limited spatial resolution of ground-based images do not allow 
the authors to study the ionization stratification at the bow shock 
\citep{Metal99}.  
A comprehensive investigation of the occurrence of bow shocks 
associated with the expansion of fast collimated outflows or 
bullets in PNe is in progress \citep{Getal08}.

\section{Summary}

Our spatio-kinematical study of IC\,4634 has revealed new structural 
components in this nebula, including what seems to be the relic of 
ancient precessing collimated ejections, a triple shell morphology, 
and a thin skin of enhanced [O~{\sc iii}]/H$\alpha$ enveloping the 
nebula and the bow shock structures at the tip of the collimated 
outflows.  
Furthermore, the \emph{HST} images and ground-based echelle 
spectroscopy provide a detailed view of the physical structure 
of the bow shock associated with a collimated outflow in a PN.  
Their morphology and kinematics have been successfully modeled using 
hydro-dynamical simulations in which a precessing fast collimated 
outflow interacts with nebular material.

IC\,4634 seems to have experienced a series of mildly asymmetric mass loss 
episodes that have removed the stellar envelope, yielding to the current 
fast stellar wind.  
These periods of mass loss are interspersed with two episodic ejections 
of collimated outflows that have interacted with the nebular material, 
having important consequences in the shaping of IC\,4634.  
The most recent episode of ejection of collimated outflows is 
directly responsible of the formation of the bow shock structures 
\emph{A-A$^\prime$}, \emph{B-B$^\prime$}, and \emph{C-C$^\prime$} 
within the envelope, and has played a significant role in the 
asymmetry and orientation of the inner shell.  
The probable interaction of this collimated outflow with 
circumstellar material might be linked to the formation 
of \emph{D-D$^\prime$}.  
Finally, the oldest episode of ejection of collimated outflows formed 
\emph{H} and \emph{Hd} in the outermost regions.  
The two ejections of collimated outflows took place at different times 
during the PN formation.  
The large misalignment between the ejection axes of the collimated 
outflows giving raise to \emph{A-A$^\prime$}, \emph{B-B$^\prime$}, 
and \emph{C-C$^\prime$}, on one hand, and \emph{H} and \emph{Hd}, on 
the other, imply that the collimating source had experienced important 
changes between the ejections.

\acknowledgments

The data presented here have been taken using ALFOSC, which is owned 
by the Instituto de Astrof\'{\i}sica de Andaluc\'{\i}a (IAA) and 
operated at the Nordic Optical Telescope under agreement between IAA 
and the NBIfAFG of the Astronomical Observatory of Copenhagen.  

Part of this work was supported by the Spanish MCyT projects number 
AYA~2002-00376 and AYA~2005-01495 cofunded by FEDER funds.  
The work of A.R.\ was supported by the Spanish MCyT grants 
AYA~2005-08523-C03-01 and AYA~2005-08013-C03-01 cofunded by FEDER funds. 
PFV and ACR acknowledge support from CONACyT (Mexico) grant
46628-F, and DGAPA-UNAM grants IN108207 and IN100407.   
The work of PFV and ACR is also supported by the ``Macroproyecto
de Tecnolog\'\i as para la Universidad de la Informaci\'on y la
Computaci\'on'' (Secretar\'\i a de Desarrollo Institucional de la UNAM,
Programa Transdisciplinario en Investigaci\'on y Desarrollo
para Facultades y Escuelas, Unidad de Apoyo a la Investigaci\'on en
Facultades y Escuelas).
PFV and ACR thank to Enrique Palacios and Mart\'\i n Cruz (ICN) for
maintaining and supporting our Linux servers and their assistance 
provided.  
LO, RV, and GB acknowledge kind support from CONACyT (Mexico) grant 
45848 and DGAPA-PAPIIT-UNAM grant IN111903.

We would like to dedicate this paper in memory of our colleague 
and friend, Dr.\ Hugo Schwarz, who recently passed away.




\clearpage




\begin{thebibliography}{}

\bibitem[Aller \& Czyzak(1983)]{AC83} 
Aller, L.~H., \& Czyzak, S.~J.\ 1983, \apjs, 51, 211 
\bibitem[Balick(1987)]{B87}
Balick, B.\ 1987, \aj, 94, 671
\bibitem[Balick(2004)]{B04} 
Balick, B.\ 2004, \aj, 127, 2262 
\bibitem[Balick et al.(1998)]{Betal98}
Balick, B., Alexander, J., Hajian, A.~R., Terzian, Y., Perinotto, M., \& 
Patriarchi, P.\ 1998, \aj, 116, 360
\bibitem[Balick \& Hajian(2004)]{BH04} 
Balick, B., \& Hajian, A.~R.\ 2004, \aj, 127, 2269
\bibitem[Balick et al.(1994)]{Betal94}
Balick, B., Perinotto, M., Maccioni, A., Terzian, Y., \& Hajian, A.~R.\ 
1994, \apj, 424, 800
\bibitem[B\"ohm \& Solf(1985)]{BS85}
 B\"ohm, K.-H. \& Solf, J. \ 1985, \apj, 294, 533
\bibitem[Cliffe et al.(1995)]{Cetal95} 
Cliffe, J.~A., Frank, A., Livio, M., \& Jones, T.~W.\ 1995, \apjl, 447, L49 
\bibitem[Corradi et al.(1996)]{Cetal96} 
Corradi, R.\ L.\ M., Guerrero, M.\ A., Manchado, A., \& Mampaso, A.\ 1996, 
New Astr, 2, 461
\bibitem[de Freitas Pacheco et al.(1992)]{dFPetal92} 
de Freitas Pacheco, J.~A., Maciel, W.~J., \& Costa, R.~D.~D.\ 1992, 
\aap, 261, 579 
\bibitem[Dopita(1997)]{D97}
Dopita, M.~A.  \ 1997, \apj, 485, L41
\bibitem[Durand et al.(1998)]{DAZ98} 
Durand, S., Acker, A., \& Zijlstra, A.\ 1998, \aaps, 132, 13 
\bibitem[Fern\'andez, Monteiro, \& Schwarz(2004)]{FMS04} 
Fern{\'a}ndez, R., Monteiro, H., \& Schwarz, H.~E.\ 2004, \apj, 603, 595 
\bibitem[Garc{\'{\i}}a-Arredondo \& Frank(2004)]{GF04} 
Garc{\'{\i}}a-Arredondo, F., \& Frank, A.\ 2004, \apj, 600, 992 
\bibitem[Garc\'{\i}a-Segura \& L\'opez(2000)]{G-SL00}
Garc{\'{\i}}a-Segura, G., \& L{\'o}pez, J.~A.\ 2000, \apj, 544, 336 
\bibitem[Gon\c calves, Corradi, \& Mampaso(2001)]{GCM01}
Gon\c calves, D.R., Corradi, R.L.M., \& Mampaso, A.\ 2001, \apj, 547, 302
\bibitem[Gruendl et al.(2000)]{Getal00} 
Gruendl, R.~A., Chu, Y.-H., Dunne, B.~C., \& Points, S.~D.\ 2000, \aj, 
120, 2670 
\bibitem[Guerrero et al.(2008)]{Getal08}
Guerrero, M.A., Medina, J.J., Luridiana, V., Miranda, L.F., Riera, A., \& 
Vel\'azquez, P.F.\ 2008, in preparation
\bibitem[Guerrero et al.(2004)]{GMC04}
Guerrero, M.~A., Miranda, L.~F., \& Chu, Y.-H.\ 2004, Asymmetrical 
Planetary Nebulae III: Winds, Structure and the Thunderbird, 313, 30 
\bibitem[Hajian et al.(1997)]{Hetal97} 
Hajian, A.~R., Balick, B., Terzian, Y., \& Perinotto, M.\ 1997, \apj, 487, 304 
\bibitem[Hartigan et al.(1987)]{HRH87}
Hartigan, P., Raymond, J. \& Hartmann, L. \ 1987, \apj, 316, 323
\bibitem[Hyung et al.(1999)]{HAF99} 
Hyung, S., Aller, L.~H., \& Feibelman, W.~A.\ 1999, \apj, 525, 294 
\bibitem[Johnson et al.(2006)]{Jetal06} 
Johnson, M.~D., Levitt, J.~S., Henry, R.~B.~C., \& Kwitter, K.~B.\ 2006, 
Planetary Nebulae in our Galaxy and Beyond, 234, 439 
\bibitem[Kwok et al.(1978)]{KPF78}
Kwok, S., Purton, C.~R., \& Fitzgerald, P.~M.\ 1978, \apjl, 219, L125 
\bibitem[Lee \& Sahai(2003)]{LS03}
Lee, C.-F., \& Sahai, R.\ 2003, \apj, 586, 319 
\bibitem[M$^{\rm c}$Keith et al.(1993)]{MKetal93} 
M$^{\rm c}$Keith, C.~D., Garc\'{\i}a-L\'opez, R.~J., Rebolo, R., Barnett, 
E.~W., Beckman, J.~E., Mart\'{\i}n, E.~L., \& Trapero, J.\ 1993, \aap, 
273, 331 
\bibitem[Medina et al.(2007)]{Metal07}
Medina, J.J., Guerrero, M.A., Luridiana, V., Miranda, L.F., Riera, A., \& 
Vel\'azque, P.F.\ 2007, proceedings of the Asymmetrical Planetary Nebulae 
IV Conference, eds.\ R.L.M.\ Corradi, A.\ Manchado and N.\ Soker, in press
\bibitem[Mendoza(1983)]{M83}
Mendoza, C.\ 1983, in Planetary Nebulae, IAU Symp., 103, 143
\bibitem[Miranda, Guerrero, \& Torrelles(1999)]{MGT99} 
Miranda, L.~F., Guerrero, M.~A., \& Torrelles, J.~M.\ 1999, \aj, 117, 1421 
\bibitem[Miranda \& Solf(1992)]{MS92} 
Miranda, L.~F., \& Solf, J.\ 1992, \aap, 260, 397 
\bibitem[Miranda et al.(1999)]{Metal99}
Miranda, L.~F., V\'azquez, R., Corradi, R.L.M., Guerrero, M.~A., L\'opez, 
J.A., \& Torrelles, J.~M.\ 1999, \apj, 520, 714
\bibitem[O'Dell et al.(2002)]{ODetal02} 
O'Dell, C.~R., Balick, B., Hajian, A.~R., Henney, W.~J., \& Burkert, A.\ 
2002, \aj, 123, 3329 
\bibitem[Patriarchi \& Perinotto(1991)]{PP91} 
Patriarchi, P., \& Perinotto, M.\ 1991, \aaps, 91, 325
\bibitem[Peimbert(1978)]{P78} 
Peimbert, M.\ 1978, IAU Symp.~ 76: Planetary Nebulae, 76, 215 
\bibitem[Perinotto et al.(2004)]{Petal04}
Perinotto, M., Patriarchi, P., Balick, B., \& Corradi, R.~L.~M.\ 2004, 
\aap, 422, 963 
\bibitem[Raga et al.(1990)]{Retal90} 
Raga, A.~C., Binette, L., Canto, J., \& Calvet, N.\ 1990, \apj, 364, 601 
\bibitem[Raga \& B\"ohm(1985)]{RB85}
Raga, A.C. \& B\"ohm, K.-H. \ 1985, \apjs, 58, 201 
\bibitem[Raga \& B\"ohm(1986)]{RB86}
Raga, A.C. \& B\"ohm, K.-H. \ 1986, \apj, 308, 829
\bibitem[Raga et al.(2002)]{Retal02}
Raga, A.C., de Gouveia Dal Pino, E.M., Noriega-Crespo, A., Mininni, P.D.,
\& Vel\'azquez, P.F. 2002, \aap, 392, 267 
\bibitem[Raga, Navarro-Gonz\'alez, \& Villagran-Muniz(2000)]{RNV00}
Raga, A.C., Navarro-Gonz\'alez, R., \& Villagran-Muniz, M.\ 2000, 
Revista Mexicana de Astr.\ y Astrof., 36, 67
\bibitem[Raga et al.(2007)]{Retal07} 
Raga, A.C., Riera, A., Mellema, G., Esquivel, A. \& Vel\'azquez, P.F. \ 2007, 
 \aap (submitted)
\bibitem[Riera \& Raga (2007)]{RR07}
Riera, A. \& Raga, A.C. \ 2007, proceedings of the Asymmetrical Planetary 
Nebulae IV Conference, eds.\ R.L.M.\ Corradi, A.\ Manchado and N.\ Soker, 
in press 
\bibitem[Rijkhorst, Mellema, \& Icke(2005)]{RMI05}
Rijkhorst, E.-J., Mellema, G., \& Icke, V.\ 2005, \aap, 444, 849
\bibitem[Sahai(2000)]{S00} 
Sahai, R.\ 2000, \apjl, 537, L43  
\bibitem[Savage \& Mathis(1979)]{SM79}  
Savage, B.D., \& Mathis, J.S.\ 1979, \araa, 17, 23
\bibitem[Schwarz(1993)]{S93} 
Schwarz, H.\ E.\ 1993, 
Mass Loss on the AGB and Beyond, Second ESO/CTIO Workshop, 
eds.\ H.\ E.\ Schwarz (ESO; Garching), 223
\bibitem[Soker \& Bisker(2006)]{SB06} 
Soker, N., \& Bisker, G.\ 2006, \mnras, 369, 1115 
\bibitem[Stanghellini, Corradi, \& Schwarz(1993)]{SCS93} 
Stanghellini, L., Corradi, R.~L.~M., \& Schwarz, H.~E.\ 1993, \aap, 279, 521 
\bibitem[Toledano et al.(2003)]{Tetal03} 
Toledano, O., Riesgo-Tirado, H., L\'opez, J.~A., Meaburn, J., Bryce, M., 
\& Holloway, A.~J.\ 2003, IAU Symposium, 209, 543 
\bibitem[Van Leer(1982)]{VL82}
Van Leer, B. 1982, ICASE Report Nos. 82-30 
\bibitem[Vel\'azquez, Riera \& Raga(2004)]{VRR04}
Vel\'azquez, P.~F., Riera, A. \& Raga, A.~C. \ 2004, \aap, 419, 991 
\end{thebibliography}
\end{document}